\documentclass[12pt]{article}
\pdfoutput=1
\usepackage{yfonts}
\usepackage{color}
\usepackage{mhchem}
\usepackage{xcolor}
\usepackage{cite}
\usepackage{hyperref}
\hypersetup{colorlinks=true,linkcolor=red,anchorcolor=black,citecolor=green}
\usepackage[toc,page]{appendix}
\usepackage{amsfonts}
\usepackage{bbold}
\usepackage{textcomp}
\usepackage[DIV13]{typearea}
\usepackage{amsmath, amsthm, amssymb, mathtools,empheq,latexsym,dsfont}
\usepackage{bbm}
\usepackage{slashed, simplewick}
\usepackage[utf8]{inputenc}
\usepackage{graphicx,placeins}
\usepackage{makeidx}
\usepackage[font=small,labelfont=bf]{caption}
\usepackage{nicefrac}
\usepackage{subfigure}
\usepackage{array, bigdelim,multirow,multicol}
\usepackage[integrals]{wasysym}
\usepackage{fancybox}
\usepackage{bm}
\usepackage{float}
\usepackage{rotating}
\usepackage{colortbl}
\usepackage{booktabs}
\usepackage[top=2cm,textwidth=16.6cm,textheight=22.75cm]{geometry}
\usepackage{doi}
\usepackage{fancybox}
\usepackage[compat=1.0.0]{tikz-feynman}
\usepackage{tikz}
\usepackage{empheq}
\usepackage{longtable}
\allowdisplaybreaks
\newcommand{\ignore}[1]{}

\usepackage{braket}
\usepackage{makecell}
\usepackage{diagbox}
\usepackage{enumitem}  

\graphicspath{{immagini/}}

\definecolor{Gray}{gray}{0.92}
\DeclareMathAlphabet{\mathscr}{OT1}{pzc}{m}{it}

\renewcommand{\arg}{{\rm Arg}}

\newcommand{\be}{\begin{equation}}
\newcommand{\ee}{\end{equation}}
\newcommand{\bea}{\begin{eqnarray}}
\newcommand{\eea}{\end{eqnarray}}

\makeatletter
\@addtoreset{equation}{section}
\makeatother

\title{\bf Froggatt-Nielsen mechanism with non-invertible symmetry}

\begin{document}

\title{
\begin{flushright}
\begin{minipage}{0.2\linewidth}
\normalsize
\end{minipage}
\end{flushright}
{\Large \bf Froggatt-Nielsen mechanism with non-invertible symmetry
\\[2mm]}
\date{}
\author{
Gui-Jun~Ding$^{a,b}$\footnote{E-mail: {\tt
dinggj@ustc.edu.cn}},  \
Stephen~F.~King$^{c}$\footnote{E-mail: {\tt king@soton.ac.uk}}, \
Bu-Yao Qu$^{a}$\footnote{E-mail: {\tt
qubuyao@mail.ustc.edu.cn}(corresponding author)},  \
\\*[20pt]
\centerline{
\begin{minipage}{\linewidth}
\begin{center}
$^a${\it \small Department of Modern Physics,  and Anhui Center for fundamental sciences in theoretical physics,\\
University of Science and Technology of China, Hefei, Anhui 230026, China}\\[2mm]
$^b${\it \small  College of Physics, Guizhou University, Guiyang 550025, China}\\[2mm]
$^c${\it \small Physics and Astronomy, University of Southampton, Southampton, SO17 1BJ, U.K.}
\end{center}
\end{minipage}}
\\[10mm]}}

\maketitle
\thispagestyle{empty}

\begin{abstract}
We formulate the Froggatt-Nielsen mechanism in the framework of non-invertible symmetries obtained by $\mathbb{Z}_2$ gauging of $\mathbb{Z}_N$ or $U(1)$. While the non-invertible selection rules can constrain the texture-zero structure of fermion mass matrices, they do not by themselves determine the magnitudes of the allowed couplings. Introducing a flavon field allows the fusion rules to fix the leading powers of the small parameter $\varepsilon=\langle S\rangle/\Lambda$, thereby determining both the texture zeros and the orders of magnitude of the nonzero entries up to $\mathcal{O}(1)$ coefficients. We have constructed explicit quark and lepton models and study their predictions for the fermion mass and mixing hierarchies through analytical estimates and numerical analysis. In particular, a benchmark model with nine texture zeros in the quark sector can well accommodate the quark masses and mixing parameters with non-Hermitian Yukawa matrices. Its lepton sector admits a minimal type-I seesaw realization with normal neutrino mass ordering and one massless neutrino. These results provide a common symmetry framework for fermion mass hierarchies and texture zeros.
\end{abstract}

\clearpage

\section{\label{sec:introd} Introduction}

The observed hierarchical patterns of fermion masses and mixing parameters constitute one of the long-standing puzzles of the Standard Model (SM). In the SM, the masses of the quarks and charged leptons are generated through their Yukawa interactions with the SM Higgs doublet, and the observed mass hierarchy reflects the vastly different strengths of these interactions~\cite{ParticleDataGroup:2026mpi}. The structure of the quark Yukawa matrices further determines the quark mixing angles and the CP-violating phase of the Cabibbo--Kobayashi--Maskawa (CKM) matrix. Consequently, the Yukawa couplings of the quarks and charged leptons span several orders of magnitude, and the quark mixing angles also exhibit a pronounced hierarchical pattern. Since the entries of the Yukawa matrices are free parameters in the SM, their peculiar hierarchical structure remains unexplained. This flavor puzzle motivates the search for a dynamical origin of the fermion mass and mixing hierarchies.

The Froggatt-Nielsen (FN) mechanism~\cite{Froggatt:1978nt} provides an elegant framework for explaining the observed hierarchy of fermion Yukawa couplings in terms of a spontaneously broken $U(1)_{FN}$ flavor symmetry. The fermions of different generations are assigned different $U(1)_{FN}$ charges, which can forbid the corresponding Yukawa interactions at the renormalizable level. The flavor symmetry is spontaneously broken at a high scale $\Lambda$ by the vacuum expectation value (VEV) of a new scalar field $S$, referred to as the flavon. The flavon is a singlet under the SM gauge group but carries a nonzero $U(1)_{FN}$ charge. Higher-dimensional operators involving appropriate powers of $S$ can then generate the effective Yukawa couplings after $U(1)_{FN}$ breaking. These couplings are suppressed by powers of $\langle S\rangle/\Lambda$, with the power determined by the flavor charges of the Higgs and fermion fields. The observed fermion mass hierarchies can thus be naturally realized through different powers of the small ratio $\langle S\rangle/\Lambda$. In a similar fashion, the discrete cyclic $\mathbb{Z}_N$ symmetry instead of the continuous $U(1)_{FN}$ symmetry can be exploited to generate hierarchical Yukawa couplings.

A complementary approach to the flavor puzzle is to constrain the structure of the fermion mass matrices through texture zeros~\cite{Weinberg:1977hb,Fritzsch:1977za,Wilczek:1977uh,Frampton:2002yf,Xing:2002ta}. Such zeros, defined in a specified flavor basis, reduce the number of independent parameters and can imply relations among masses, mixing angles and CP-violating phases. Flavor symmetries provide a way to enforce these restrictions rather than imposing them as phenomenological assumptions. Nevertheless, specifying the vanishing entries alone generally leaves the magnitudes of the allowed entries unexplained. Combining symmetry-enforced texture zeros with FN suppressions is therefore a natural strategy for addressing both the structure and the hierarchy of the fermion mass matrices.

Recent developments in generalized global symmetries have enlarged the possibilities for constructing flavor theories~\cite{Gaiotto:2014kfa,Cordova:2022ruw}. Of particular interest are non-invertible symmetries, whose topological symmetry operators obey fusion rules more general than group multiplication and need not possess inverses, see Refs.~\cite{Cordova:2022ruw,Schafer-Nameki:2023jdn,Brennan:2023mmt,Bhardwaj:2023kri,Shao:2023gho} for recent review on non-invertible symmetries and references therein. In the constructions considered here, such fusion rules arise from gauging automorphisms of an underlying Abelian symmetry. For example, $\mathbb{Z}_2$ gauging identifies an element of $\mathbb{Z}_N$ with its inverse, organizing the symmetry data into equivalence classes. The resulting fusion rules determine which combinations of field assignments can contain the identity class and hence which interactions are allowed. This supplies selection rules that can be used to constrain flavor structures.

Applications of these ideas to flavor physics have shown that non-invertible selection rules can generate texture zeros in quark Yukawa matrices~\cite{Kobayashi:2024cvp,Kobayashi:2025znw}, charged lepton and light neutrino mass matrices~\cite{Kobayashi:2025ldi,Kobayashi:2025cwx,Chen:2025awz,Qu:2026omn}, and the Dirac and Majorana mass matrices of the minimal seesaw~\cite{Jiang:2025psz}. Suitable quark textures have also been used to explore axion-less solutions to the strong CP problem~\cite{Liang:2025dkm,Kobayashi:2025thd,Kobayashi:2025rpx}. In these texture-based constructions, however, the non-invertible selection rules determine the allowed and forbidden entries without fixing the magnitudes of the allowed couplings. The nonzero entries remain free parameters whose values must be chosen to reproduce the observed masses and mixing. Consequently, a successful texture does not by itself explain the hierarchy among the nonzero Yukawa couplings. This limitation motivates the present work: by incorporating the FN mechanism into the non-invertible symmetry framework, we seek to determine both the texture-zero structure and the parametric sizes of the allowed entries through symmetry-controlled powers of a small expansion parameter.

In this work, we formulate the FN mechanism for non-invertible symmetries obtained by $\mathbb{Z}_2$ gauging of $\mathbb{Z}_N$ and $U(1)$. The leading suppression of each allowed coupling is determined by the smallest number of flavon insertions compatible with the fusion rules. If no such insertion makes the operator invariant, the corresponding entry remains forbidden within the effective operator construction. The field assignments thus determine both texture zeros and the leading powers of $\varepsilon$, while leaving $\mathcal{O}(1)$ coefficients undetermined. Since $S$ and $S^\dagger$ have the same assignment in these $\mathbb{Z}_2$ constructions, a real flavon can be used in the non-supersymmetric models. We also identify cases in which the resulting suppressions coincide with those of conventional Abelian models, clarifying the role of the Higgs and flavon assignments.

We illustrate this framework with models based on $\mathbb{Z}_2$ gauging of $\mathbb{Z}_N$ and three models based on $\mathbb{Z}_2$ gauging of $U(1)$. Analytical estimates and numerical scans characterize the fermion mass and mixing hierarchies obtained with $\mathcal{O}(1)$ coefficients. Particular emphasis is placed on the benchmark  Model III, in which the non-invertible symmetry enforces its texture-zero pattern, while the FN mechanism specifies the order of magnitude of every allowed mass-matrix element. The independent, non-Hermitian quark Yukawa entries permit an excellent fit to the quark masses and CKM parameters. In the lepton sector, we employ the type-I seesaw mechanism with two right-handed neutrinos yielding one massless light neutrino as in the minimal seesaw~\cite{King:1999mb,Frampton:2002qc}. We study its lepton observables and neutrino mass predictions, distinguishing the consequences of the texture from those that depend on the fitted coefficients. The scalar vacuum structure and an extension to $\mathbb{Z}_3$ gauging of $\mathbb{Z}_N$ provide further perspectives on the framework.

The remainder of this paper is organized as follows. In section~\ref{sec:framework-FN}, we formulate the FN mechanism for non-invertible symmetries obtained by $\mathbb{Z}_2$ gauging of $\mathbb{Z}_N$ and $U(1)$, deriving the suppression powers and texture zeros of Yukawa couplings and Majorana mass terms which are determined by the selection rules. Section~\ref{sec:model-ZN-gauging} presents a model based on $\mathbb{Z}_2$ gauging of $\mathbb{Z}_{12}$ and examines its quark and lepton mass hierarchies and mixing parameters through analytical estimates and numerical scans. In section~\ref{sec:model-U(1)-gauging}, we construct three models based on $\mathbb{Z}_2$ gauging of $U(1)$ and study their flavor phenomenology. Particular attention is paid to the benchmark Model III, in which the non-invertible selection rules and the FN mechanism determine both the texture zeros and the orders of magnitude of the nonzero mass-matrix elements, allowing numerical fits in the quark and lepton sectors. Section~\ref{sec:conclusion} summarizes our results and discusses directions for further study. Appendix~\ref{app:VEV-H-S} analyzes the scalar potential and the conditions for a stable vacuum with nonzero Higgs and flavon VEVs, and discusses possible cosmological implications of the flavon sector. Appendix~\ref{app:models-Z2-gauging-ZN14-ZN36} provides additional models based on $\mathbb{Z}_2$ gauging of $\mathbb{Z}_{36}$ symmetry, together with their fermion mass textures and numerical results. Finally, appendix~\ref{app:Z3-gauhing-Z18} extends the construction to $\mathbb{Z}_3$ gauging of $\mathbb{Z}_N$ and illustrates it with a $\mathbb{Z}_{18}$ model that generates a hierarchical right-handed neutrino mass spectrum.

\section{\label{sec:framework-FN}Froggatt-Nielsen mechanism from non-invertible symmetry}

The Froggatt-Nielsen (FN) mechanism is a popular approach to explain the mass hierarchies of quarks and charged leptons~\cite{Froggatt:1978nt}. An additional $U(1)_{FN}$ flavor symmetry is introduced in this approach, and the quarks and leptons are assumed to carry various $U(1)_{FN}$ charges. The $U(1)_{FN}$ is spontaneously broken by the vacuum expectation value (VEV) of a new scalar field $S$ which is neutral under the SM gauge symmetry and it caries a negative unit of the $U(1)_{FN}$ charge by convention. The SM Yukawa couplings are not compatible with the $U(1)_{FN}$ symmetry except the top quark whose mass is of the same order as the electroweak scale. One has to include some power of $S$ to compensate the FN charges of quark and lepton fields, so that  the effective Yukawa couplings of light quarks and charged leptons are suppressed by powers of the small ration $\langle S\rangle/\Lambda$, where $\Lambda$ denotes the cut-off scale. In the following, we shall formulate the FN mechanism in the framework of the $\mathbb{Z}_2$ gauging of $\mathbb{Z}_N$ or $U(1)$ symmetry in the following. The quark Yukawa interactions invariant under the non-invertible symmetry can be generally written as
\begin{eqnarray}
\label{eq:Yukawa-quark}- \mathcal{L} = c_u^{ij} \overline{Q}_i \widetilde{H} u_j \left( \dfrac{S}{\Lambda} \right)^{n_{ij}} + c_d^{ij} \overline{Q}_i H d_j \left( \dfrac{S}{\Lambda} \right)^{m_{ij}} +\text{h.c.}
\end{eqnarray}
where $c_u^{ij}$ and $c_d^{ij}$ with $i, j=1, 2, 3$ are dimensionless couplings, assumed to be $\mathcal{O}(1)$. The scalar field $S$ is charged under the $\mathbb{Z}_2$ gauging of $\mathbb{Z}_N$ or $U(1)$ symmetry. Since $S$ and $S^{\dagger}$ transform in the same way under $\mathbb{Z}_2$ gauging of $\mathbb{Z}_N$ or $U(1)$ symmetry, we assume $S$ is a real scalar field for simplicity. In comparison with the conventional $\mathbb{Z}_N$ or $U(1)$ flavor symmetry, the flavon $S$ is a complex scalar. The power indices $m_{ij}$ and $n_{ij}$ are non-negative integers and they are determined by the invariance under the $\mathbb{Z}_2$ gauging of $\mathbb{Z}_N$ or $U(1)$ symmetry. Moreover, each term in Eq.~\eqref{eq:Yukawa-quark} can be added to any powers of $S^2$ which is invariant under the concerned non-invertible symmetry. As shown in the Appendix~\ref{app:VEV-H-S}, the scalar $S$ can develop stable VEV $\langle S\rangle$ in certain parameter space for the scalar potential fulfilling both SM gauge symmetry and non-invertible symmetry. Then the effective Yukawa couplings between the SM Higgs and fermions can be generated,  
\begin{eqnarray}
- \mathcal{L} = Y_u^{ij} \overline{Q}_i \widetilde{H} u_j + Y_d^{ij} \overline{Q}_i H d_j + \text{h.c.}\,,
\end{eqnarray}
where 
\begin{equation}
Y^{ij}_u=c^{ij}_u\varepsilon^{n_{ij}}\,,\quad Y^{ij}_d=c^{ij}_d\varepsilon^{m_{ij}}\,,\quad \varepsilon\equiv\dfrac{\langle S \rangle}{\Lambda}\,.
\end{equation}
We see that the Yukawa couplings $Y_u^{ij}$ and $Y_d^{ij}$ are suppressed by $\varepsilon^{n_{ij}}$ and $\varepsilon^{m_{ij}}$ respectively, the resulting hierarchical structure can help to explain the quark mass hierarchies and the small quark mixing angles.


\subsection{\label{sub:FN-ZN-gauging} $\mathbb{Z}_2$ gauging of $\mathbb{Z}_N$ symmetry}

Let $g$ be a generator of $\mathbb{Z}_N$ symmetry with $g^N = e$, the $\mathbb{Z}_2$ symmetry is a subgroup of the automorphism group of $\mathbb{Z}_N$, and it is generated by the element $r$ which fulfills~\cite{Bartsch:2022mpm,Kobayashi:2024cvp} 
\begin{eqnarray}
r g r^{-1} = g^{-1}\,,~~~~r^2=e\,,
\end{eqnarray}
where $e$ stands for the unit element. Using this automorphism, one defines equivalence classes
\begin{eqnarray}
[g^k] = \{ g^k, g^{-k} \}~~~~ k = 0, 1, \cdots \, \left\lfloor \dfrac{N}{2} \right\rfloor \,,
\end{eqnarray}
where $\lfloor x \rfloor$ denotes the floor function, i.e., the greatest integer less than or equal to $x$. In the following, we use the $[k]$ as a shorthand notation for the conjugacy class $[g^k]$, for simplicity. The fusion rules among these classes take the form
\begin{eqnarray}
\label{eq:fusion-rule-zngagugez2} [k][k'] = [k+k'] + [k-k'] \,,
\end{eqnarray}
which is precisely the fusion algebra obtained by $\mathbb{Z}_2$ gauging of $\mathbb{Z}_N$. To be more explicit, the equivalence classes for $\mathbb{Z}_2$ gauging of $\mathbb{Z}_{2n}$ are
\begin{eqnarray}
[0] = \{ e \} \,,~~~ [n] = \{ g^n \}\,,~~~ [k] = \{ g^k, g^{-k} \}~~~ \text{for } 1\leq k < n\,.
\end{eqnarray}
The fusion rules for $\mathbb{Z}_2$ gauging of $\mathbb{Z}_{2n}$ are given by
\begin{eqnarray}
\nonumber [0] \times [k] &=& [k]\,,~~ [n] \times [k] = [n - k] \,,~~ [0] \times [n] = [n] \,,~~ [0] \times [0] = [n] \times [n] = [0] \,, \\
\label{eq:FR-Z2n} [k]\times [l] &=& \begin{cases}
2[0] + [k+l] \,, ~~ & k=l \text{ and } k+l \neq n  \\
2[n] + [k-l] \,, ~~ & k\neq l \text{ and } k+l = n  \\
2[0] + 2[n] \,, ~~ & k=l \text{ and } k+l = n  \\
[k+l] + [k-l] \,,~~ & \text{other cases} 
\end{cases} 
\end{eqnarray}
where $1\leq k, l < n$. Analogously the equivalence classes for $\mathbb{Z}_2$ gauging of $\mathbb{Z}_{2n+1}$ symmetry are
\begin{eqnarray}
[0] = \{ e \} \,,~~~ [k] = \{ g^k, g^{-k} \}~~~ \text{for } 1\leq k \leq n\,.
\end{eqnarray}
The fusion rules for $\mathbb{Z}_2$ gauging of $\mathbb{Z}_{2n+1}$ are given by
\begin{eqnarray}
 [0] \times [k] &=& [k]\,, ~~~ [k] \times [l] = \begin{cases}
2[0] + [k+l] \,, ~~ & k=l  \\
[k+l] + [k-l] \,,~~ & \text{other cases} 
\end{cases}
\end{eqnarray}
where $1\leq k,l \leq n$.

Let us consider some special cases of $\mathbb{Z}_2$ gauging of $\mathbb{Z}_N$ models. Let the scalar fields $H$ and $S$ be assigned to the equivalence classes $H\sim [0]$ and $S\sim [1]$. The assignments of the quark fields are $Q \sim ([k_{Q_1}], [k_{Q_2}], [k_{Q_3}])$, $u \sim ([k_{u_1}], [k_{u_2}], [k_{u_3}])$ and $d \sim ([k_{d_1}], [k_{d_2}], [k_{d_3}])$, where all the integers satisfies $0 \leq k_{\psi} \leq N/2$. Then we have
\begin{eqnarray}
\nonumber n_{ij} &=& \min \left\{ |k_{Q_i} - k_{u_j}|, k_{Q_i} + k_{u_j}, N - k_{Q_i} - k_{u_j} \right\} \,, \\
m_{ij} &=& \min \left\{ |k_{Q_i} - k_{d_j}|, k_{Q_i} + k_{d_j}, N - k_{Q_i} - k_{d_j} \right\} \,.
\end{eqnarray}
Since $0\leq k_{Q_i}, k_{u_j}, k_{d_j}\leq N/2$, we have
\begin{eqnarray}
n_{ij} = |k_{Q_i} - k_{u_j}| ~~~\text{and}~~~ m_{ij} = |k_{Q_i} - k_{d_j}|\,. \label{eq:nij-mij-H0}
\end{eqnarray}
We can find the exponent of the small expansion parameter $\varepsilon\equiv\langle S\rangle/\Lambda$ only depends on the mismatch among the fermion charges. Hence we will not be able to distinguish between non-invertible symmetry and $\mathbb{Z}_N$ symmetry in this case for non-supersymmetric model. In supersymmetric models, the Yukawa superpotential must be a holomorphic function of the chiral superfields. If a non-invertible symmetry is imposed, the indices $n_{ij}$ and $m_{ij}$ are unchanged and follow Eq.~\eqref{eq:nij-mij-H0}. However, in the case of a traditional $\mathbb{Z}_N$ flavor symmetry where $S$ is a chiral superfield carrying negative unit charge of $\mathbb{Z}_N$, the indices are determined by $n_{ij}=(k_{Q_i} + k_{u_j})\;(\text{mod}~N)$ and $m_{ij} = (k_{Q_i}+k_{d_j})\;(\text{mod}~N)$, with the corresponding Yukawa couplings vanishing when the sums of charges are negative. However, if a $\bar{S}$ chiral superfield carrying positive unit charge of $\mathbb{Z}_N$ is introduced simultaneously~\cite{Ibanez:1994ig}, in order to cancel anomalies, the difference between the predictions of SUSY and non-SUSY models for both usual and non-invertible symmetry disappears.

We now turn to the alternative scenario in which the Higgs field is assigned to a nontrivial class rather than the identity class $[0]$. In this case, we take $H\sim [k_H]\neq[0]$ and $S\sim[1]$. We reorder $k_{Q_i}$, $k_{u_j}$ and $k_H$ as $0\leq k_1 \leq k_2 \leq k_3 \leq N/2$, where $k_1$ denotes the smallest among $k_{Q_i}$, $k_{u_j}$ and $k_H$, $k_3$ the largest, and $k_2$ the remaining one. Then using the fusion rule, we find that the power $n_{ij}$ of the leading order contribution is given by
\begin{eqnarray}
\nonumber n_{ij} &=& \min \left\{ k_3 + k_2 + k_1, k_3 + k_2 - k_1, k_3 - k_2 + k_1, |k_3 - k_2 - k_1|, |k_3 + k_2 + k_1 - N|,\right.  \\
&&  \left.  N - (k_3 + k_2 - k_1), N - (k_3 - k_2 + k_1), N - |k_3 - k_2 - k_1|  \right\}\,. 
\end{eqnarray}
The value of $n_{ij}$ depends on whether $2(k_1 + k_2)$ is smaller or larger than $N$. Explicitly, we find
\begin{eqnarray}
\label{eq:nij} n_{ij} = \begin{cases}
|k_3 - k_2 - k_1|\,,~~~&\text{for }~ 2(k_1 + k_2) \leq N  \\
|k_3 + k_2 + k_1 - N|\,,~~~&\text{for }~ 2(k_1 + k_2) \geq N
\end{cases}
\end{eqnarray}
In the remaining scenarios, where $k_H\neq 0$ and $k_S\neq 1$, the exponent $n_{ij}$ generally does not have a simple analytical expression.
If $k_S$ is coprime to $N$, one can redefine the charges to map the system to the case $k_S = 1$. However, when $k_S$ is not coprime to $N$, such a redefinition is not possible. Let the greatest common divisor of $k_S$ and $N$ be $d>1$. In this case, the exponent $n_{ij}$ can be taken as the smallest integer in the following sequence
\begin{eqnarray}
\dfrac{| \pm k_{Q_i} \pm  k_{u_j} \pm k_H |}{k_S}, \dfrac{| \pm k_{Q_i} \pm  k_{u_j} \pm k_H + N |}{k_S}, \dfrac{| \pm k_{Q_i} \pm  k_{u_j} \pm k_H + 2N |}{k_S}, \cdots \,.
\end{eqnarray}
If none of the elements among $k_{Q_i} \pm  k_{u_j} \pm k_H$ is multiple of $d$, then the above sequence contains no integers, and the corresponding Yukawa term is forbidden.

For the Majorana neutrinos, the Lagrangian for the mass term is given by
\begin{eqnarray}
\mathcal{L} \ni -\frac{1}{2}M c_N^{ij} \overline{\nu^c_{Ri}} \nu_{Rj} \left( \dfrac{S}{\Lambda} \right)^{n^N_{ij}} \,,
\end{eqnarray} 
where $M$ is the overall Majorana neutrino mass scale, and $c^{ij}_N=c^{ji}_N$ are dimensionless coupling constants of order one. Let the neutrino be assigned to $\nu^c_{Ri} \sim [k_{\nu^c_{Ri}}]$, then the exponent $n^N_{ij}=n^N_{ji}$ is given by
\begin{eqnarray}
n^N_{ij}={\substack{\text{min} \\ \text{integer candidates}}}\left\{\frac{|\pm k_{\nu^c_{Ri}}\pm k_{\nu^c_{Rj}}|}{k_S}, \frac{|\pm k_{\nu^c_{Ri}}\pm k_{\nu^c_{Rj}}+N|}{k_S}, \frac{|\pm k_{\nu^c_{Ri}}\pm k_{\nu^c_{Rj}}+2N|}{k_S},\ldots\right\} \,.~~~
\end{eqnarray}
For the case of unit flavon label with $k_S=1$, it follows that 
\begin{eqnarray}
\label{eq:nMij}n^N_{ij} = |k_{\nu^c_{Ri}} - k_{\nu^c_{Rj}}| \,.
\end{eqnarray}
This differs from the conventional $\mathbb{Z}_N$ case, where the corresponding exponent is simply $k_{\nu^c_{Ri}} + k_{\nu^c_{Rj}}$.

\subsection{\label{sub:FN-U(1)-gauging} $\mathbb{Z}_2$ gauging of $U(1)$ symmetry}

When $N$ is very large, the non-invertible $\mathbb{Z}_N$ symmetry would mimick a non-invertible $U(1)$ symmetry. The group $U(1)$ has an automorphism group $\mathbb{Z}_2$ whose generator is denoted as $r$ and $U(1) \rtimes \mathbb{Z}_2 \cong O(2)$ as shown in~\cite{Lee00}. A generic element of $U(1)$ can be generally written as $g=e^{i\theta}$, where $\theta$ is a continuously varying real quantity.  
The action of $r$ on $g$ is $r g r^{-1}=g^{-1}$. Under this automorphism, we see that $g$ and its inverse $g^{-1}$ belong to the same equivalence class as well as $[k]=\{e^{ik\theta}, e^{-ik\theta}\}$, where $k$ id non-negative integer. Without loss of generality, the charge $k$ of the $\mathbb{Z}_2$ gauging of $U(1)$ symmetry can be taken to be positive.
If the field $S$ carries unit $U(1)$ charge, from Eq.~\eqref{eq:nij} and taking the limit $N\rightarrow \infty$, the exponents $n_{ij}$ are given by  
\begin{eqnarray}
n_{ij} = |k_3 - k_2 - k_1| \,.
\end{eqnarray}
If the field $S$ carries a non-negative integer $U(1)$ charge $k_S$ and quarks and Higgs field also carry non-negative integer charges, then the power index $n_{ij}$ is given by 
\begin{eqnarray}
\text{min}\left\{ \left| \dfrac{k_{Q_i} + k_{u_j} + k_{H} }{k_S} \right| , \left| \dfrac{k_{Q_i} + k_{u_j} - k_{H} }{k_S} \right| , \left| \dfrac{k_{Q_i} - k_{u_j} + k_{H} }{k_S} \right|, \left| \dfrac{k_{Q_i} - k_{u_j} - k_{H} }{k_S} \right| \right\}
\end{eqnarray} 
If none of them is an integer, the corresponding Yukawa coupling is forbidden. 

For Majorana neutrinos, analogously the power index is the smallest integer among the following values,
\begin{eqnarray}
\text{min}\left\{ \left| \dfrac{k_{\nu^c_{Ri}} + k_{\nu^c_{Rj}}}{k_S} \right|, \left| \dfrac{k_{\nu^c_{Ri}} - k_{\nu^c_{Rj}}}{k_S} \right| \right\} \,.
\end{eqnarray}
If none of them is an integer, the corresponding Majorana mass term is forbidden. For $S$ carrying the unit $U(1)$ charge, the power index for the Majorana neutrino mass term is given by
\begin{eqnarray}
n^N_{ij} = |k_{\nu^c_{Ri}} - k_{\nu^c_{Rj}}| \,.
\end{eqnarray}

\section{\label{sec:model-ZN-gauging} Model based on $\mathbb{Z}_2$ gauging of $\mathbb{Z}_N$ symmetry }

We consider the $\mathbb{Z}_2$ gauging $\mathbb{Z}_{12}$ symmetry, the equivalence classes are 
\begin{eqnarray}
[0] = \{ e \}\,,~~~ [6] = \{g^6 \}\,,~~~  [k] = \{ g^k, g^{-k} \} ~~~\text{for } 1\leq k \leq 5 \,.
\end{eqnarray}
The scalar field are assigned to be $H\sim [0]$ and $S\sim [1]$ respectively. The quark fields are assigned to be
\begin{eqnarray}
(Q_1, Q_2, Q_3) \sim ([0], [1], [3]) \,,~~~ (u_1, u_2, u_3) \sim ([6], [4], [3]) \,,~~~ (d_1, d_2, d_3) \sim ([6], [5], [5])
\end{eqnarray}
With these assignment the mass matrices have the order of magnitude entries
\begin{eqnarray}
\label{eq:Mu-Md-N12} M_u \sim \begin{pmatrix}
\varepsilon^6 ~&~ \varepsilon^4 ~&~ \varepsilon^3   \\
\varepsilon^5 ~&~ \varepsilon^3 ~&~ \varepsilon^2   \\
\varepsilon^3 ~&~ \varepsilon ~&~ 1
\end{pmatrix} v_H \,,~~~
M_d \sim \begin{pmatrix}
\varepsilon^6 ~&~ \varepsilon^5 ~&~ \varepsilon^5   \\
\varepsilon^5 ~&~ \varepsilon^4 ~&~ \varepsilon^4   \\
\varepsilon^3 ~&~ \varepsilon^2 ~&~ \varepsilon^2
\end{pmatrix} v_H \,.
\end{eqnarray}
Because of the hierarchical structure, it is possible to analytically derive some approximate results for the singular values and the rotations to the mass basis. One can also perturbatively diagonalize the Yukawa matrices, and
obtain the following estimates for the quark masses and CKM matrix elements:
\begin{eqnarray}
\nonumber&& m_u\sim\varepsilon^{6}v_H\,,~~m_c\sim\varepsilon^{3}v_H\,,~~m_t\sim v_H\,,\\
\nonumber && m_d\sim \varepsilon^{6}v_H\,,~~m_s\sim \varepsilon^{4}v_H\,,~~m_b\sim \varepsilon^{2}v_H\,,\\
\label{eq:N12_quark_para}&&V_{us}\sim\varepsilon\,,~~V_{ub}\sim\varepsilon^3\,,~~V_{cb}\sim\varepsilon^2\,,
\end{eqnarray}
where coefficients of order one are neglected. 
Analogously the lepton fields are assigned to transform under the non-invertible symmetry as follow,
\begin{eqnarray}
\ell \sim ([6], [5], [5])\,,~~~ E_R \sim ([0], [1], [3])\,,~~~ \nu_R \sim ([0], [1], [1])
\end{eqnarray}
The charged lepton mass and neutrino mass matrices are
\begin{eqnarray}
\nonumber M_e &\sim & \begin{pmatrix}
\varepsilon^6 ~&~ \varepsilon^5 ~&~ \varepsilon^3   \\
\varepsilon^5 ~&~ \varepsilon^4 ~&~ \varepsilon^2   \\
\varepsilon^5 ~&~ \varepsilon^4 ~&~ \varepsilon^2
\end{pmatrix} v_H \,,~~~
M_D \sim  \begin{pmatrix}
\varepsilon^6 ~&~ \varepsilon^5 ~&~ \varepsilon^5   \\
\varepsilon^5 ~&~ \varepsilon^4 ~&~ \varepsilon^4   \\
\varepsilon^5 ~&~ \varepsilon^4 ~&~ \varepsilon^4
\end{pmatrix} v_H \,,~~~   
M_N \sim \begin{pmatrix}
1 ~&~ \varepsilon ~&~ \varepsilon   \\
\varepsilon ~&~ 1 ~&~ 1   \\
\varepsilon ~&~ 1 ~&~ 1
\end{pmatrix} M \,,~~~   \\
M_N^{-1} &\sim & \begin{pmatrix}
1 ~&~ \varepsilon ~&~ \varepsilon   \\
\varepsilon ~&~ 1 ~&~ 1   \\
\varepsilon ~&~ 1 ~&~ 1
\end{pmatrix} M^{-1} \,,~~~
M_\nu \sim \begin{pmatrix}
\varepsilon^{10} ~&~ \varepsilon^9 ~&~ \varepsilon^9   \\
\varepsilon^9 ~&~ \varepsilon^8 ~&~ \varepsilon^8   \\
\varepsilon^9 ~&~ \varepsilon^8 ~&~ \varepsilon^8
\end{pmatrix} \dfrac{v_H^2}{M} \,. \label{eq:lepton-MM-N12}
\end{eqnarray}
The diagonal entries of the right-handed neutrino mass matrix are generally unsuppressed by powers of $\varepsilon$ in the framework of $\mathbb{Z}_2$ gauging of $\mathbb{Z}_N$ symmetry, consequently the masses of the three right-handed neutrinos are of the same order of magnitude. The $2\times2$ subblock in the lower-right corner of the light neutrino mass matrix $M_{\nu}$ does not exhibit any particular structure or symmetry, and it can naturally gives to a large $\theta_{23}$ together with small $\theta_{13}\sim\varepsilon$. The order of magnitudes predictions for the charged lepton masses and neutrino mixing angels parameters,
\begin{eqnarray}
\nonumber&& m_e\sim\varepsilon^6 v_H\,,~~ m_\mu\sim\varepsilon^4 v_H\,,~~m_{\tau}\sim\varepsilon^2 v_H\,, \\
\nonumber &&\sin\theta^{\ell}_{12}\sim\varepsilon\,,~~\sin\theta^{\ell}_{13}\sim\varepsilon\,,~~\sin\theta^{\ell}_{23}\sim1 \,, \\
&& m_1 \sim \varepsilon^{12} \dfrac{v_H^2}{M} \,,~~ m_2 \sim \varepsilon^{8} \dfrac{v_H^2}{M}\,,~~ m_3 \sim \varepsilon^{8} \dfrac{v_H^2}{M}\,. \label{eq:lepton-mass-mixing-N12}
\end{eqnarray}
Both $\theta_{12}$ and $\theta_{13}$ are predicted to be of the order $\varepsilon$, consequently fine-tuning of couplings are required in order to explain the experimentally measured large value of solar mixing angle~\cite{Esteban:2024eli}.

We randomly draw some large number of $\mathcal{O}(1)$ complex coefficient matrices $Y_{u,d}$ and calculate the quark masses and mixing parameters for each instance of $Y_{u,d}$ as a function of $\varepsilon$. The absolute values of the $Y_{u,d}^{ij}$ are taken to be uniformly distribute in $[1/3, 3]$ on a logarithmic scale. The phases are chosen to be uniformly distributed in $[0,2\pi)$. For a given choice of $\varepsilon$, we can then calculate the observable with the maximum fractional deviation from its SM value~\cite{Cornella:2023zme}:
\begin{eqnarray}
\Xi_{\rm quark} = \max_i \left\{ \left| \dfrac{O_i}{\mu_i} \right| , \left| \dfrac{\mu_i}{O_i} \right|  \right\}
\end{eqnarray}
where $O_i$ are the model predictions for the observables and $\mu_i$ denote the central values of the corresponding quantities obtained from experimental data, as shown in table~\ref{tab:quark_data}. We minimize $\Xi_{\rm quark}$ with respect to $\varepsilon$ with range over $[0.1,0.3]$ for each instance of the $Y_{u,d}$ coefficient matrices.

\begin{figure}[hptb!]
\centering
\includegraphics[width=0.45\textwidth]{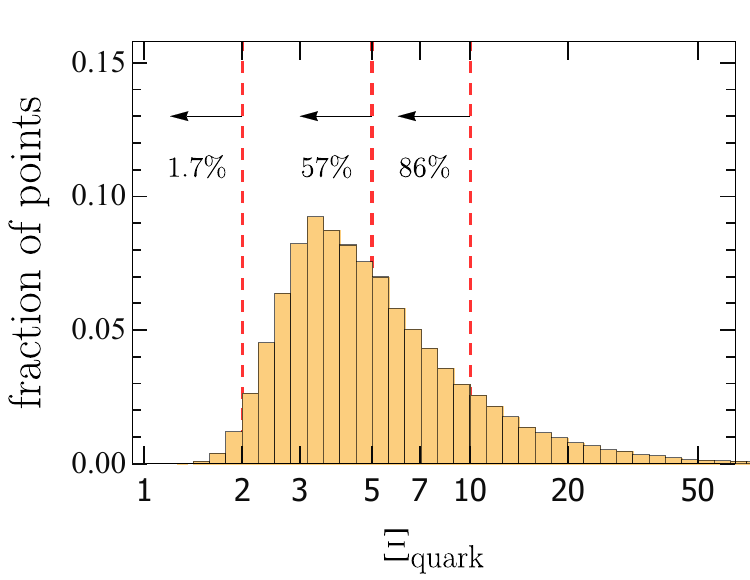}
\includegraphics[width=0.45\textwidth]{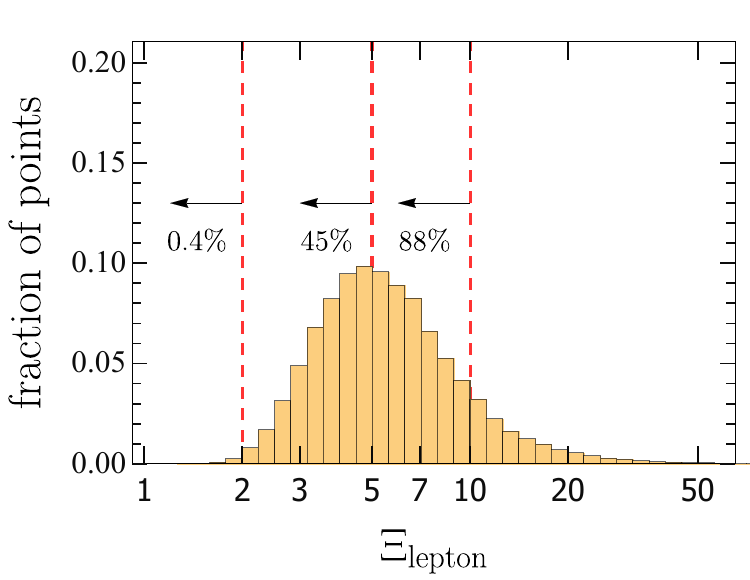}
\caption{\label{fig:N12_Xi}The probability distribution function for $\Xi_{\rm quark}$ and $\Xi_{\rm lepton}$ for the model based on $\mathbb{Z}_2$ gauging of $\mathbb{Z}_{12}$ symmetry, respectively. The total fraction of coefficient matrices that result in $\Xi_{\rm quark(lepton)} < 2, 5$ and $10$ are shown respectively. The random $\mathcal{O}(1)$ coefficients are sampled from the range $[1/3,3]$.  }
\end{figure}

To assess how readily the texture in Eq.~\eqref{eq:Mu-Md-N12} reproduces the observed quark hierarchies, we examine the distribution of the optimized $\Xi_{\rm quark}$ in the left panel of figure~\ref{fig:N12_Xi}. The histogram peaks around $\Xi_{\rm quark}\sim3$--$4$, and the fractions of sampled coefficient matrices yielding $\Xi_{\rm quark}<2$, $5$ and $10$ are $1.7\%$, $57\%$ and $86\%$, respectively. Thus, more than half of the samples reproduce all observables entering this measure within a factor of five of their central values, whereas agreement within a factor of two is much less common. These fractions characterize the performance of the texture under the specified sampling prescription; they are not statistical confidence levels or fractions of points satisfying the experimental error bars. The scan supports an explanation of the overall flavor hierarchies, while a precision fit requires selecting suitable coefficients.

\begin{figure}[hptb!]
\centering
\includegraphics[width=6.5in]{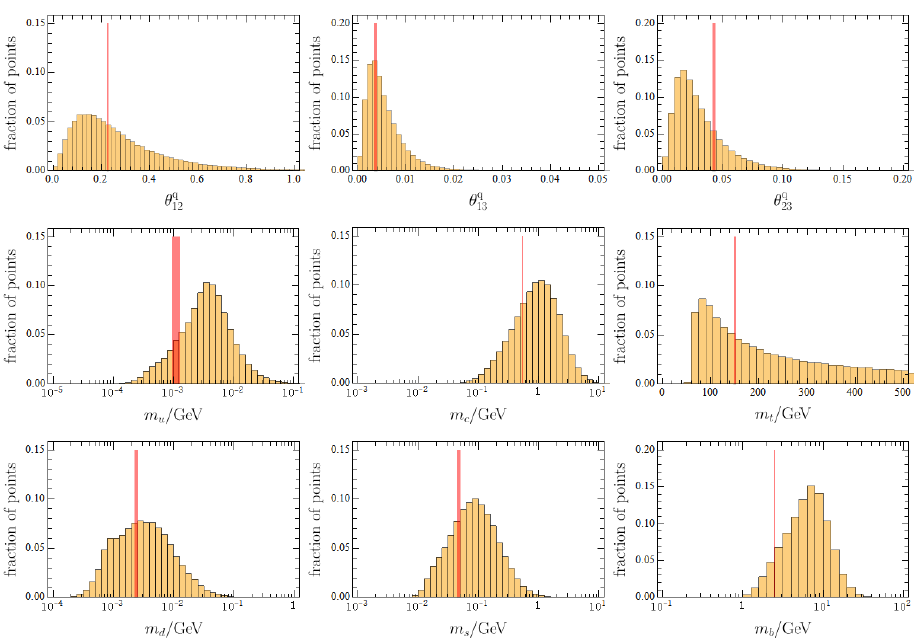}
\caption{\label{fig:N12_quark_obs}Distributions of the quark masses and mixing parameters for the model based on $\mathbb{Z}_2$ gauging of $\mathbb{Z}_{12}$ symmetry. The red shadows mark the SM values for the observables with $1\sigma$ region at $\mu = 1\,{\rm TeV}$. }
\end{figure}

Figure~\ref{fig:N12_quark_obs} shows how this behavior is distributed among the individual observables. The $\theta^q_{13}$ and $m_d$ distributions have appreciable weight near the experimental bands, whereas the peaks for $m_u$, $m_c$ and $m_b$ lie above their measured values; $m_s$ shows a smaller upward shift. The $\theta^q_{12}$ and $\theta^q_{23}$ distributions peak below their experimental values but have broad tails extending across the preferred regions. These trends are consistent with the competing requirements on the common expansion parameter. The quark scan favors a typical value $\varepsilon\simeq0.15$, for which the leading estimates in Eq.~\eqref{eq:N12_quark_para} give $|V_{us}|\sim0.15$, $|V_{cb}|\sim0.023$ and $|V_{ub}|\sim0.0034$. The first two require mild enhancements of the order one coefficients to reach the observed values, while the third already has the appropriate size. Increasing $\varepsilon$ would raise these mixing angles but also enhance the light quark masses through the higher powers in Eq.~\eqref{eq:N12_quark_para}, making the upward mass shifts more difficult to compensate. The unsuppressed top Yukawa coupling instead gives $m_t\sim v_H$, with its broad distribution mainly reflecting the $\mathcal{O}(1)$ coefficients. The plots demonstrate that the FN power counting successfully accounts for the observed flavor hierarchies. Given the large number of free parameters, the model has sufficient flexibility to reproduce the experimental observables through suitable parameter choices. Moreover, the large parameter space tends to result in relatively weak correlations among the physical observables.
Regarding the CP-violation, the CP phase in the CKM matrix exhibits no significant preference because of its dependence on randomly distributed phases of the couplings. We therefore do not display its distribution in figure.

\begin{table}[t!]
\centering
\begin{tabular}{| c | c | } \hline \hline
Observable & Central value and $1\sigma$ error  \\ \hline
$m_u / {\rm GeV} $ & $0.00107 \pm 0.000024 $   \\
$m_c / {\rm GeV} $ & $0.541 \pm 0.0087 $   \\
$m_t / {\rm GeV} $ & $149.9 \pm 0.75 $   \\
$m_d / {\rm GeV} $ & $0.00235 \pm 0.000035$   \\
$m_s / {\rm GeV} $ & $0.0466 \pm 0.00052 $   \\
$m_b / {\rm GeV} $ & $2.437_{-0.05}^{+0.016} $   \\ \hline
$\delta^q_{CP}/ {}^\circ$ & $65.26 \pm 1.32$  \\
$\theta^q_{12}$ & $0.22704 \pm 0.00082$   \\
$\theta^q_{13}$ & $0.00377 \pm 0.00008$  \\
$\theta^q_{23}$ & $0.04275\pm 0.00042$  \\ \hline\hline
\end{tabular}
\caption{\label{tab:quark_data}The central values and the $1\sigma$ errors of the mass and mixing angles and CP violation phases in quark sector, where we adopt the renormalization scale $\mu=1\,{\rm TeV}$ in the $\overline{\rm MS}$ scheme~\cite{Antusch:2025fpm}.  }
\end{table}

For the lepton sector, the observables we consider are the charged lepton masses, the PMNS matrix elements $|U_{12}|$, $|U_{13}|$, $|U_{21}|$, $|U_{23}|$, $|U_{31}|$ and $|U_{32}|$, the neutrino mass square differences, $\Delta m_{21}^2$ and $\Delta m_{31}^2$. The measured values and uncertainties for these parameters are listed in table~\ref{tab:lepton_data}. For each observable $\mathcal{O}_i$, the fractional derivation from the experimental value is defined as~\cite{Cornella:2024jaw}
\begin{eqnarray}
\Xi_{\rm lepton} = \max_i \delta_i \,,~~~\text{with}~~~
\delta_{i} = \begin{cases}
\dfrac{\mu_i^{\rm min}}{\mathcal{O}_i}  \,,~~~& \mathcal{O}_i < \mu_i^{\rm min} \\
1\,,~~~&   \mu_i^{\rm min} \leq \mathcal{O}_i \leq \mu_i^{\rm max} \\
\dfrac{\mathcal{O}_i}{\mu_i^{\rm max}}  \,,~~~& \mathcal{O}_i > \mu_i^{\rm max}
\end{cases}
\end{eqnarray}
Here $\mu_i^{\rm min}$ and $\mu_i^{\rm max}$ are the experimental lower and upper bounds for the observable $\mathcal{O}_i$. 
Similar to the quark sector, we generate random coefficient matrices of $Y_{e,D}$ and $M_N$. Then we adjust $\varepsilon$ and the right handed neutrino mass scale $M$ to minimize the $\Xi_{\rm lepton}$.

\begin{figure}[t!]
\centering
\includegraphics[width=6.5in]{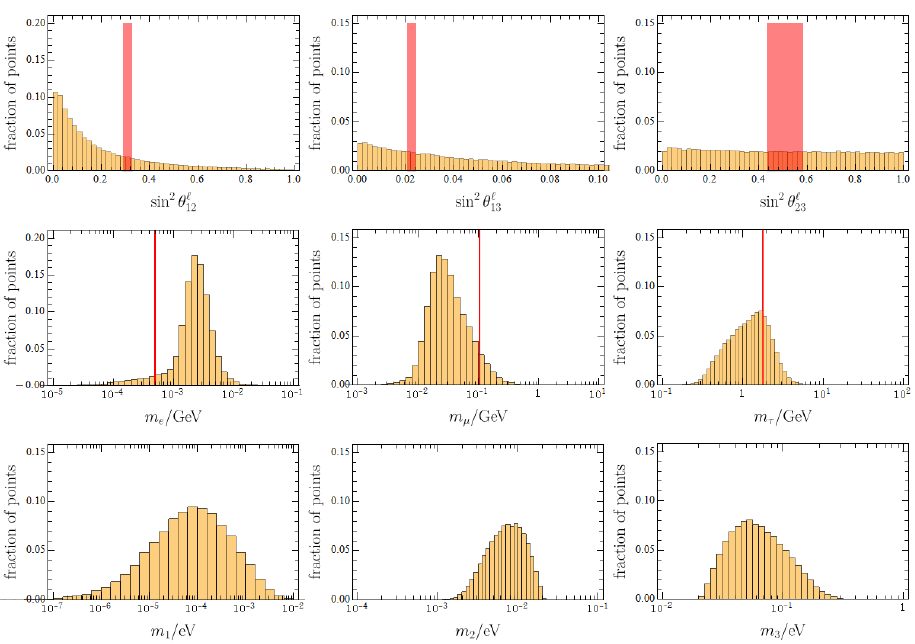}
\caption{\label{fig:N12_lepton_obs}Distributions of the lepton masses and mixing parameters for the model based on $\mathbb{Z}_2$ gauging of $\mathbb{Z}_{12}$ symmetry. The red shadows mark the SM values for the observables with $1\sigma$ region at $\mu = 1\,{\rm TeV}$. }
\end{figure}

The corresponding lepton results are displayed in the right panel of figure~\ref{fig:N12_Xi} and in figure~\ref{fig:N12_lepton_obs}. The $\Xi_{\rm lepton}$ distribution peaks around $5$, with $0.4\%$, $45\%$ and $88\%$ of the samples satisfying $\Xi_{\rm lepton}<2$, $5$ and $10$, respectively. Under the definition above, $\Xi_{\rm lepton}=1$ means that all fitted observables lie within their adopted experimental intervals, whereas $\Xi_{\rm lepton}<2$ only requires agreement within a factor of two of the interval boundaries. The small fraction below two shows that obtaining a close simultaneous description of the lepton observables is more restrictive than reproducing their broad hierarchy. The quark and lepton measures also use different experimental reference prescriptions, so their numerical values should not be compared as identical goodness-of-fit statistics.

For the lepton sector, the characteristic expansion parameter is $\varepsilon\sim0.2$. The charged lepton texture then gives $m_e/m_\mu\sim m_\mu/m_\tau\sim\varepsilon^2\simeq0.04$, whereas the ratios inferred from table~\ref{tab:lepton_data} are approximately $0.0047$ and $0.059$. The muon-to-tau hierarchy is therefore readily accommodated, but the electron-to-muon ratio needs an additional suppression from the coefficients or their combinations. This feature is reflected in figure~\ref{fig:N12_lepton_obs}: the electron mass distribution peaks at a few MeV, above the experimental value, while the muon distribution favors a few tens of MeV, below its measured value. The tau distribution has substantial weight near the experimental band. Lowering $\varepsilon$ alone cannot remove both charged lepton offsets, since it suppresses the muon mass as well as the electron mass. A closer fit consequently relies on the relative coefficients and the structure of the singular values of $M_e$.

The mixing distributions reveal a complementary limitation. The leading estimates $\sin\theta_{12}^\ell\sim\sin\theta_{13}^\ell\sim\varepsilon$ imply squared mixing angles of order $0.04$ for $\varepsilon\sim0.2$. This is compatible in order of magnitude with the reactor angle, whose measured region lies within the broad $\sin^2\theta_{13}^\ell$ distribution. The solar angle requires an enhancement over the expansion parameter. Its histogram is concentrated at small values, with the experimental band on the falling tail, indicating that a large solar angle requires suitable coefficient ratios and interference between the charged lepton and neutrino rotations. By contrast, the $\sin^2\theta_{23}^\ell$ distribution is broad and approximately flat, consistent with the absence of a suppression between the second and third generations. Large atmospheric mixing is thus readily obtained, but neither maximal mixing nor a particular octant is selected.

The neutrino mass distributions exhibit a pronounced hierarchy: $m_1$ is typically concentrated around $10^{-4}\,{\rm eV}$, with a broad spread, while $m_2$ lies around $10^{-2}\,{\rm eV}$ and $m_3$ around a few $10^{-2}\,{\rm eV}$. This behavior is qualitatively consistent with the strong suppression of $m_1$ relative to $m_{2,3}$ in the analytical estimates above. The lightest neutrino is small but is not forced to vanish in this three-right-handed-neutrino model. Since $m_2$ and $m_3$ carry the same leading FN power, their observed splitting depends on the coefficients, while the fitted scale $M$ sets the overall neutrino mass scale. Taken together, the figures show that this texture captures the broad charged lepton and neutrino hierarchies and allows large atmospheric mixing, with the electron mass and solar mixing angle presenting the clearest challenges to a generic fit. Finally, the quoted quark and lepton values of $\varepsilon$ come from separate optimizations. As $\varepsilon=\langle S\rangle/\Lambda$ is common to the model, establishing a simultaneous description of both sectors requires a joint fit at a single value of $\varepsilon$.

\begin{table}[t!]
\centering
\begin{tabular}{| c | c | } \hline \hline
Observable & Experimental value  \\ \hline
$m_e / {\rm MeV} $ & $0.4910 \pm 0.00010 $   \\
$m_\mu / {\rm MeV} $ & $103.4\pm 0.023$   \\
$m_\tau / {\rm MeV} $ & $1756.81 \pm 0.35 $   \\
$\Delta m_{21}^2 / (10^{-5}\,{\rm eV}^2) $ &  $7.236\rightarrow 7.823$    \\
$\Delta m_{31}^2 / (10^{-3}\,{\rm eV}^2) $ &  $2.450\rightarrow 2.576$   \\ \hline
$|U_{12}|$ & $0.532 \rightarrow 0.568$  \\
$|U_{13}|$ & $0.144 \rightarrow 0.156$  \\
$|U_{21}|$ & $0.259 \rightarrow 0.498$  \\
$|U_{23}|$ & $0.652 \rightarrow 0.756$  \\
$|U_{31}|$ & $0.280 \rightarrow 0.512$  \\
$|U_{32}|$ & $0.488 \rightarrow 0.659$  \\ \hline\hline
\end{tabular}
\caption{\label{tab:lepton_data}The central values and the $1\sigma$ errors of the charged lepton masses~\cite{Antusch:2025fpm} and the $3\sigma$ range for neutrino mass square differences and the element of lepton mixing matrix~\cite{Esteban:2024eli}, where $U_{ij}$ is element of the lepton mixing matrix. }
\end{table}

\section{\label{sec:model-U(1)-gauging} Models based on $\mathbb{Z}_2$ gauging of $U(1)$ symmetry }

In this section, we shall present three models for quarks and leptons based on the $\mathbb{Z}_2$ gauging of $U(1)$ symmetry, and the quark and lepton mass matrices are of different textures in these three models.

\subsection{\label{sub:modelI-U(1)-gauging} Model I}

The scalar field are assigned to be $H\sim [6]$ and $S\sim [1]$ respectively. The quark fields are assigned to be
\begin{eqnarray}\nonumber 
(Q_1, Q_2, Q_3) \sim ([0], [1], [15])\,,~~~ (u_1, u_2, u_3) \sim ([13], [1], [9])\,,~~~ (d_1, d_2, d_3) \sim ([12], [1], [11])
\end{eqnarray}
The mass matrices are
\begin{eqnarray}\label{eq:quark_texture_case1}
M_u \sim \begin{pmatrix}
\varepsilon^7 ~&~ \varepsilon^5 ~&~ \varepsilon^3   \\
\varepsilon^6 ~&~ \varepsilon^4 ~&~ \varepsilon^2  \\
\varepsilon^4 ~&~ \varepsilon^8 ~&~ 1
\end{pmatrix} v_H \,,~~~
M_d \sim \begin{pmatrix}
\varepsilon^6 ~&~ \varepsilon^5 ~&~ \varepsilon^5   \\
\varepsilon^5 ~&~ \varepsilon^4 ~&~ \varepsilon^4   \\
\varepsilon^3 ~&~ \varepsilon^8 ~&~ \varepsilon^2
\end{pmatrix} v_H \,.
\end{eqnarray}
The quark masses and CKM matrix elements can be estimated as follows 
\begin{eqnarray}
\nonumber&& m_u\sim\varepsilon^{7}v_H\,,~~m_c\sim\varepsilon^{4}v_H\,,~~m_t\sim v_H\,,\\
\nonumber && m_d\sim \varepsilon^{6}v_H\,,~~m_s\sim \varepsilon^{4}v_H\,,~~m_b\sim \varepsilon^{2}v_H\,,\\
&&V_{us}\sim\varepsilon\,,~~V_{ub}\sim\varepsilon^3\,,~~V_{cb}\sim\varepsilon^2\,. \label{eq:quark-mass-mixing-modelI}
\end{eqnarray}
The lepton fields are assigned to 
\begin{eqnarray}
\ell \sim ([12], [11], [1])\,,~~~ E_R \sim ([0], [1], [3])\,,~~~ \nu_R \sim ([0], [4], [8])
\end{eqnarray}
The charged lepton mass and neutrino mass matrices are
\begin{eqnarray}\nonumber 
M_e &\sim & \begin{pmatrix}
\varepsilon^6 ~&~ \varepsilon^5 ~&~ \varepsilon^3   \\
\varepsilon^5 ~&~ \varepsilon^4 ~&~ \varepsilon^2   \\
\varepsilon^5 ~&~ \varepsilon^4 ~&~ \varepsilon^2
\end{pmatrix} v_H \,,~~~
M_D \sim  \begin{pmatrix}
\varepsilon^6 ~&~ \varepsilon^2 ~&~ \varepsilon^2   \\
\varepsilon^5 ~&~ \varepsilon ~&~ \varepsilon^3   \\
\varepsilon^5 ~&~ \varepsilon ~&~ \varepsilon
\end{pmatrix} v_H \,,~~~   
M_N \sim \begin{pmatrix}
1 ~&~ \varepsilon^4 ~&~ \varepsilon^8   \\
\varepsilon^4 ~&~ 1 ~&~ \varepsilon^4   \\
\varepsilon^8 ~&~ \varepsilon^4 ~&~ 1
\end{pmatrix} M \,,~~~   \\
M_N^{-1} &\sim & \begin{pmatrix}
1 ~&~ \varepsilon^4 ~&~ \varepsilon^8   \\
\varepsilon^4 ~&~ 1 ~&~ \varepsilon^4   \\
\varepsilon^8 ~&~ \varepsilon^4 ~&~ 1
\end{pmatrix} M^{-1} \,,~~~
M_\nu \sim \begin{pmatrix}
\varepsilon^4 ~&~ \varepsilon^3 ~&~ \varepsilon^3   \\
\varepsilon^3 ~&~ \varepsilon^2 ~&~ \varepsilon^2   \\
\varepsilon^3 ~&~ \varepsilon^2 ~&~ \varepsilon^2
\end{pmatrix} \dfrac{v_H^2}{M} \,. \label{eq:Me_Mnu_case1}
\end{eqnarray}
The order of magnitudes predictions for the charged lepton masses and neutrino mixing angels parameters,
\begin{eqnarray}
\nonumber&& m_e\sim\varepsilon^6 v_H\,,~~ m_\mu\sim\varepsilon^4 v_H\,,~~m_{\tau}\sim\varepsilon^2 v_H\,, \\
\nonumber &&\sin\theta^{\ell}_{12}\sim\varepsilon\,,~~\sin\theta^{\ell}_{13}\sim\varepsilon\,,~~\sin\theta^{\ell}_{23}\sim1 \,, \\
&& m_1 \sim \varepsilon^{12} \dfrac{v_H^2}{M} \,,~~ m_2 \sim \varepsilon^2 \dfrac{v_H^2}{M}\,,~~ m_3 \sim \varepsilon^2 \dfrac{v_H^2}{M}\,.
\end{eqnarray}

\begin{figure}[t!]
\centering
\includegraphics[width=0.45\textwidth]{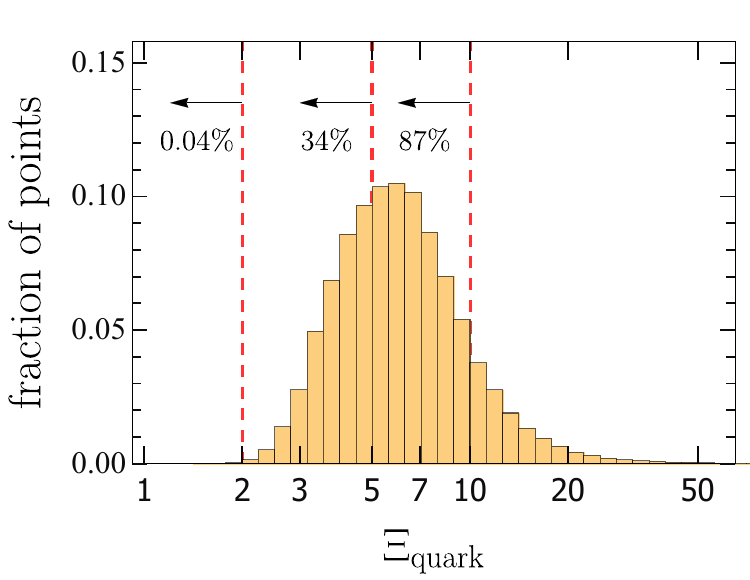}
\includegraphics[width=0.45\textwidth]{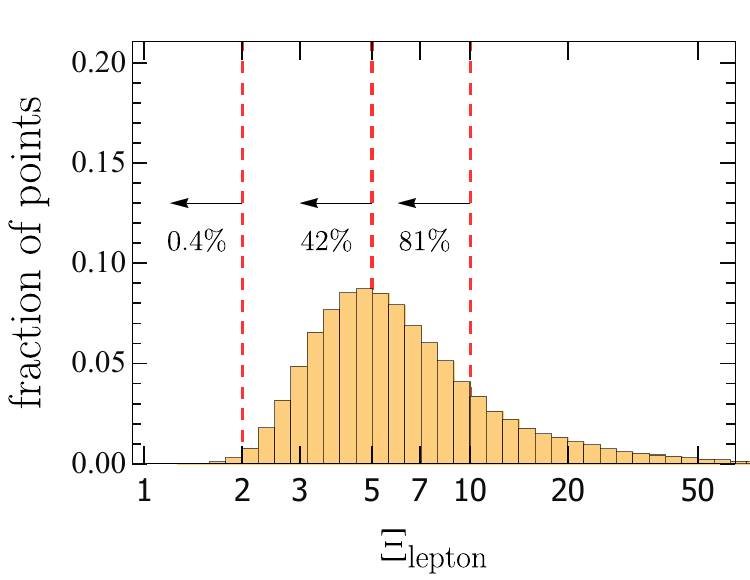}
\caption{\label{fig:case1_Xi}A probability distribution function for $\Xi_{\rm quark}$ and $\Xi_{\rm lepton}$ for the Model I. The total fraction of coefficient matrices that result in $\Xi_{\rm quark} < 2, 5$ and $10$ are shown respectively. The random $\mathcal{O}(1)$ coefficients are sampled from the range $[1/3,3]$.  }
\end{figure}

\begin{figure}[t!]
\centering
\includegraphics[width=6.5in]{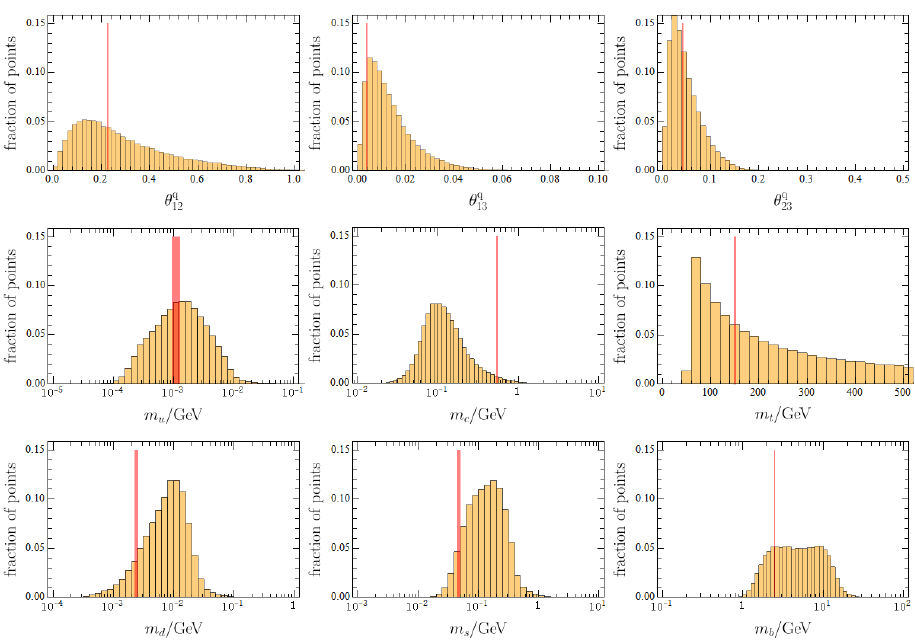}
\caption{\label{fig:case1_quark_obs}Distributions of the quark masses and mixing parameters for the Model I. The red shadows mark the SM values for the observables with $1\sigma$ region at $\mu = 1\,{\rm TeV}$. }
\end{figure}

We perform a numerical analysis for the benchmark Model I using the same procedure as in section~\ref{sec:model-ZN-gauging}, including the random sampling of $\mathcal{O}(1)$ coefficients and the criteria for assessing agreement with the observed fermion masses and mixing parameters.
In figure~\ref{fig:case1_Xi}, we summarize the overall agreement through the distributions of $\Xi_{\rm quark}$ and $\Xi_{\rm lepton}$.
Both distributions are concentrated at values of order a few. Consequently, for the majority of the sampled points, the predicted observables remain within one order of magnitude of their corresponding experimental values, characterized by $\Xi<10$.
This supports the role of the FN powers in setting the overall flavor hierarchies across the sampled coefficient space. The smaller fractions at $\Xi<2$ indicate that a more precise simultaneous description of experimental data selects particular combinations of coefficients. 
The distributions of the quark masses and CKM mixing parameters in Model I are shown in figure~\ref{fig:case1_quark_obs}. Each observable is concentrated within a factor of a few of its corresponding SM value. In particular, the upper panels exhibit the expected FN hierarchy $|V_{us}|\sim\varepsilon$, $|V_{cb}|\sim\varepsilon^2$, and $|V_{ub}|\sim\varepsilon^3$, with the experimentally allowed regions well within the populated ranges. The quark masses illustrate both the successes and limitations of the FN power counting. Since the up quark mass carries one additional power of $\varepsilon$ relative to the down quark mass, i.e., $m_u/m_d\sim\varepsilon$, the ordering $m_u<m_d$ arises naturally from the FN structure itself rather than solely from a hierarchy among the $\mathcal{O}(1)$ coefficients. Correspondingly, the down quark distribution peaks at a higher mass scale than the up quark distribution. By contrast, both $m_c$ and $m_s$ scale as $\varepsilon^4v_H$ in Eq.~\eqref{eq:quark-mass-mixing-modelI}, so the observed hierarchy between $m_c$ and $m_s$ cannot be attributed to FN power counting alone and instead requires an appropriate hierarchy among the corresponding $\mathcal{O}(1)$ coefficients. Consistently, this is reflected in figure~\ref{fig:case1_quark_obs}, where the experimentally allowed charm quark mass lies above the peak of its distribution, whereas the strange-quark mass lies below its corresponding peak.
Raising $\varepsilon$ alone does not resolve it, since $m_c$ and $m_s$ now carry the same leading power. The model does not predict a preferred value for the quark CP-violating phase, since the phases of the individual coupling constants are assumed to be randomly distributed. 

\begin{figure}[t!]
\centering
\includegraphics[width=6.5in]{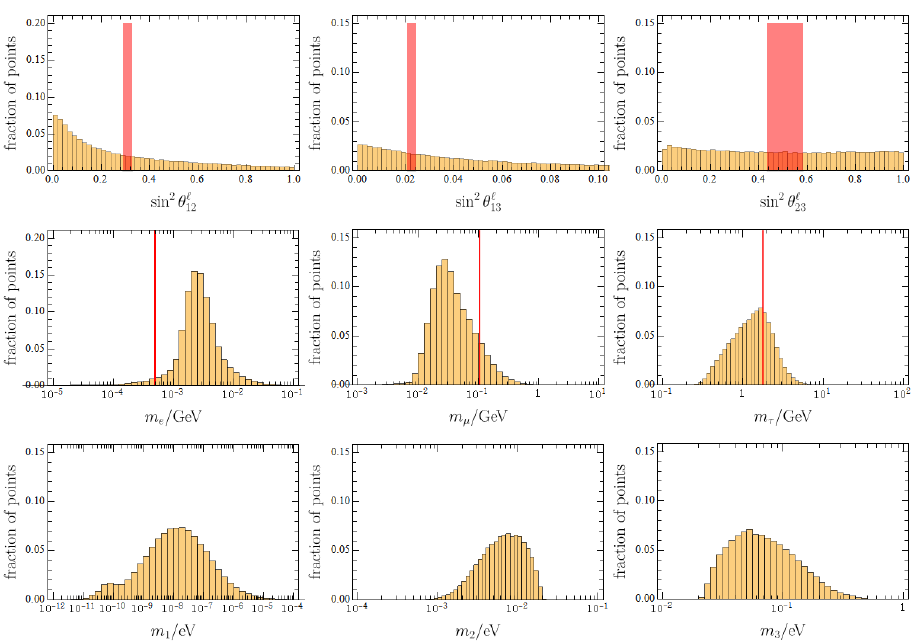}
\caption{\label{fig:case1_lepton_obs}Distributions of the lepton masses and mixing parameters for the Model I. The red shadows mark the SM values for the observables with $3\sigma$ region at $\mu = 1\,{\rm TeV}$. }
\end{figure}

In the lepton sector, the charged lepton mass matrix has the same pattern as that in Eq.~\eqref{eq:lepton-MM-N12}, and therefore the charged lepton masses exhibit the same hierarchical structure as shown in Eq.~\eqref{eq:lepton-mass-mixing-N12}. At leading order, the light-neutrino mass matrix also has the same structure as that in section~\ref{sec:model-ZN-gauging}. Consequently, the Model I and the lepton model discussed section~\ref{sec:model-ZN-gauging} exhibit the same lepton mixing pattern in Eq.~~\eqref{eq:lepton-MM-N12}.
Therefore, additional parameter tuning is still required to suppress the electron mass and to generate a large solar neutrino mixing angle.
No significant preference is observed for $\theta^\ell_{23}$, the Dirac CP-violating phase, and the Majorana phases, with the latter phases reflecting the random phases assigned to the free coefficients.
The present model and the model of section~\ref{sec:model-ZN-gauging} differ in the neutrino Yukawa matrix $M_D$ and the right handed neutrino mass matrix $M_N$.  As can be seen from Eq.~\eqref{eq:Me_Mnu_case1}, the right handed neutrino mass matrix is approximately diagonal in the Model I, while the neutrino Dirac mass matrix is dominated by its second and third columns. Consequently, the resulting light-neutrino mass matrix is approximately of rank two. The first column of the Dirac mass matrix then provides a small correction to the light-neutrino masses, giving the lightest neutrino a small but nonzero mass. The analytic estimates $m_1/m_{2,3}\sim\varepsilon^{10}$, rather than $\varepsilon^4$ in the $\mathbb{Z}_2$ gauging of $\mathbb{Z}_{12}$ model, explain the strongly displaced $m_1$ distribution in figure~\ref{fig:case1_lepton_obs}. 
Accordingly, the model predicts a normally ordered spectrum with a very light but nonzero $m_1$, typically below $0.1\,{\rm meV}$.
Moreover, the light masses $m_{2,3}$ scale as $\varepsilon^2v_H^2/M$, rather than $\varepsilon^8v_H^2/M$ in the model of section~\ref{sec:model-ZN-gauging}. For comparable coefficients and light neutrino masses, Model I therefore favors a substantially higher seesaw scale.

\subsection{\label{sub:modelII-U(1)-gauging} Model II}

The scalar field are assigned to be $H\sim [8]$ and $S\sim [3]$ respectively. The quark fields are assigned to be
\begin{eqnarray}\nonumber 
(Q_1, Q_2, Q_3) \sim ([3], [10], [4])\,,~~~ (u_1, u_2, u_3) \sim ([0], [10], [4])\,,~~~ (d_1, d_2, d_3) \sim ([3], [10], [6])
\end{eqnarray}
The mass matrices are
\begin{eqnarray}
M_u \sim \begin{pmatrix}
0 ~&~ \varepsilon^5 ~&~ \varepsilon^3   \\
\varepsilon^6 ~&~ \varepsilon^4 ~&~ \varepsilon^2  \\
\varepsilon^4 ~&~ \varepsilon^2 ~&~ 1
\end{pmatrix} v_H \,,~~~
M_d \sim \begin{pmatrix}
0 ~&~ \varepsilon^5 ~&~ 0   \\
\varepsilon^5 ~&~ \varepsilon^4 ~&~ \varepsilon^4   \\
\varepsilon^3 ~&~ \varepsilon^2 ~&~ \varepsilon^2
\end{pmatrix} v_H \,.
\end{eqnarray}
The quark masses and CKM matrix elements can be estimated as follows 
\begin{eqnarray}
\nonumber&& m_u\sim\varepsilon^{7}v_H\,,~~m_c\sim\varepsilon^{4}v_H\,,~~m_t\sim v_H\,,\\
\nonumber && m_d\sim \varepsilon^{6}v_H\,,~~m_s\sim \varepsilon^{4}v_H\,,~~m_b\sim \varepsilon^{2}v_H\,,\\
&&V_{us}\sim\varepsilon\,,~~V_{ub}\sim\varepsilon^3\,,~~V_{cb}\sim\varepsilon^2\,.
\end{eqnarray}
We find that the $(11)$ element of the down-quark mass matrix $M_d$ vanishes, while the $(12)$ and $(21)$ elements are both of order $\varepsilon^5$, although they are not exactly identical. Consequently, we obtain a GST-like relation~\cite{Gatto:1968ss},
\begin{equation}
\theta^{q}_{12}\sim\sqrt{\frac{m_d}{m_s}}\,,
\end{equation}
which relates the Cabibbo angle $\theta^{q}_{12}$ to the down- and strange-quark masses.

The lepton fields are assumed to transform under the $\mathbb{Z}_2$ gauging of $U(1)$ symmetry as follows,
\begin{eqnarray}
\ell \sim ([10], [7], [9])\,,~~~ E_R \sim ([0], [3], [7])\,,~~~ \nu_R \sim ([0], [9], [7]) \,.
\end{eqnarray}
The charged lepton mass and neutrino mass matrices are
\begin{eqnarray}\nonumber 
M_e &\sim & \begin{pmatrix}
\varepsilon^6 ~&~ \varepsilon^5 ~&~ \varepsilon^3   \\
\varepsilon^5 ~&~ \varepsilon^4 ~&~ \varepsilon^2   \\
0 ~&~ 0 ~&~ \varepsilon^2
\end{pmatrix} v_H \,,~~~
M_D \sim  \begin{pmatrix}
\varepsilon^6 ~&~ \varepsilon^3 ~&~ \varepsilon^3   \\
\varepsilon^5 ~&~ \varepsilon^2 ~&~ \varepsilon^2   \\
0 ~&~ 0 ~&~ \varepsilon^2
\end{pmatrix} v_H \,,~~~   
M_N \sim \begin{pmatrix}
1 ~&~ \varepsilon^3 ~&~ 0   \\
\varepsilon^3 ~&~ 1 ~&~ 0   \\
0 ~&~ 0 ~&~ 1
\end{pmatrix} M \,,~~~   \\
M_N^{-1} &\sim & \begin{pmatrix}
1 ~&~ \varepsilon^3 ~&~ 0   \\
\varepsilon^3 ~&~ 1 ~&~ 0   \\
0 ~&~ 0 ~&~ 1
\end{pmatrix} M^{-1} \,,~~~
M_\nu \sim \begin{pmatrix}
\varepsilon^6 ~&~ \varepsilon^5 ~&~ \varepsilon^5   \\
\varepsilon^5 ~&~ \varepsilon^4 ~&~ \varepsilon^4   \\
\varepsilon^5 ~&~ \varepsilon^4 ~&~ \varepsilon^4
\end{pmatrix} \dfrac{v_H^2}{M} \,.
\end{eqnarray}
The order of magnitudes predictions for the charged lepton masses and neutrino mixing angels parameters,
\begin{eqnarray}
\nonumber&& m_e\sim\varepsilon^6 v_H\,,~~ m_\mu\sim\varepsilon^4 v_H\,,~~m_{\tau}\sim\varepsilon^2 v_H\,, \\
\nonumber &&\sin\theta^{\ell}_{12}\sim\varepsilon\,,~~\sin\theta^{\ell}_{13}\sim\varepsilon\,,~~\sin\theta^{\ell}_{23}\sim1 \,, \\
&& m_1 \sim \varepsilon^{12} \dfrac{v_H^2}{M} \,,~~ m_2 \sim \varepsilon^4 \dfrac{v_H^2}{M}\,,~~ m_3 \sim \varepsilon^4 \dfrac{v_H^2}{M}\,.
\end{eqnarray}
It is notable that the right-handed neutrino masses are generally quasi-degenerate in the $\mathbb{Z}_2$ gauging of $\mathbb{Z}_N$ or $U(1)$ symmetry. This also has some implication for leptogenesis such as resonant leptogenesis.

\subsection{\label{sub:modelIII-U(1)-gauging} Model III}

Now, let us consider the framework of two Higgs doublet models, in which two Higgs fields $H_u$ and $H_d$ are introduced to couple to up-type quarks and down-type quarks respectively. The scalar fields are assigned to be $H_u \sim [10]$, $H_d \sim [15]$ and $S \sim [7]$. The quark fields are assigned to be
\begin{eqnarray} \nonumber 
(Q_1, Q_2, Q_3) \sim ([0], [23], [29])\,,~~~ (u_1, u_2, u_3) \sim ([29], [25], [19])\,,~~~ (d_1, d_2, d_3) \sim ([4], [27], [0])
\end{eqnarray} 
The mass matrices are
\begin{eqnarray}
M_u \sim \begin{pmatrix}
0 ~&~ \varepsilon^5 ~&~ 0   \\
\varepsilon^6 ~&~ 0 ~&~ \varepsilon^2  \\
0 ~&~ \varepsilon^2 ~&~ 1
\end{pmatrix} v_u \,,~~~
M_d \sim \begin{pmatrix}
0 ~&~ \varepsilon^6 ~&~ 0   \\
\varepsilon^6 ~&~ \varepsilon^5 ~&~ 0   \\
0 ~&~ 0 ~&~ \varepsilon^2
\end{pmatrix} v_d \,,
\end{eqnarray}
where $v_u$ and $v_d$ are the VEV of the Higgs $H_{u}$ and $H_d$ respectively. We use the conventional normalization $\sqrt{v_u^2 + v_d^2} = 174\,{\rm GeV}$ and take $v_u\simeq v_d \simeq 123\,{\rm GeV}$.

The lepton fields are assigned to 
\begin{eqnarray}
\ell \sim ([0], [23], [15])\,,~~~ E_R \sim ([11], [27], [9])\,,~~~ \nu_R \sim ([11], [19])
\end{eqnarray}
The charged lepton mass and neutrino mass matrices are
\begin{eqnarray}\nonumber 
M_e &\sim & \begin{pmatrix}
0 ~&~ \varepsilon^6 ~&~ 0   \\
\varepsilon^7 ~&~ \varepsilon^5 ~&~ 0   \\
0 ~&~ 0 ~&~ \varepsilon^3
\end{pmatrix} v_d \,,~~~ 
M_D \sim  \begin{pmatrix}
\varepsilon^3 ~&~ 0   \\
0 ~&~ \varepsilon^2   \\
\varepsilon^2 ~&~ \varepsilon^2
\end{pmatrix} v_u \,,~~~  
M_N \sim \begin{pmatrix}
1 ~&~ 0   \\
0 ~&~ 1
\end{pmatrix} M \,,~~~   \\
M_N^{-1} &\sim & \begin{pmatrix}
1 ~&~ 0   \\
0 ~&~ 1
\end{pmatrix} M^{-1} \,,~~~
M_\nu \sim \begin{pmatrix}
\varepsilon^6 ~&~ 0 ~&~ \varepsilon^5   \\
0 ~&~ \varepsilon^4 ~&~ \varepsilon^4   \\
\varepsilon^5 ~&~ \varepsilon^4 ~&~ \varepsilon^4
\end{pmatrix} \dfrac{v_u^2}{M} \,.
\end{eqnarray}

This quark texture has previously been considered in the context of non-parallel six-texture-zero Hermitian quark mass matrices~\cite{Ponce:2011qp}. It was found to be incompatible with the experimentally allowed parameter ranges, as the resulting charm-quark mass $m_c$ is significantly smaller than its measured value. 
In the present work, we relax the Hermiticity assumption and treat the parameters in the quark mass matrices as independent. The resulting texture pair is maximally restrictive while remaining compatible with the observed quark masses and mixing parameters, containing a total of nine texture zeros. The nine texture zeros in these two matrices cannot be enforced by a conventional Abelian FN symmetry without enlarging the scalar sector beyond $H_u$, $H_d$ and a single flavon $S$. Although, Abelian symmetries can realize the same zero pattern if additional scalars are introduced~\cite{Grimus:2004hf}, non-invertible symmetries provide a more economical realization.
In this model, the quark mass matrix can be parameterized as
\begin{eqnarray}\label{eq:ModelIII_Mud}
M_u \sim \begin{pmatrix}
0 ~&~ a \varepsilon^5 ~&~ 0   \\
b \varepsilon^6 ~&~ 0 ~&~ c \varepsilon^2  \\
0 ~&~ d \varepsilon^2 ~&~ e
\end{pmatrix} v_u \,,~~~
M_d \sim \begin{pmatrix}
0 ~&~ a' \varepsilon^6 ~&~ 0   \\
b' \varepsilon^6 ~&~ c' e^{i\phi}\, \varepsilon^5 ~&~ 0   \\
0 ~&~ 0 ~&~ d' \varepsilon^2
\end{pmatrix} v_d \,.
\end{eqnarray}
After suitable fields rephasing, we can take all coupling constants to be real and positive, with the CP-violating phase residing in a single coupling. We choose to place this phase in the $(2,2)$ entry $Y_d^{22}$. Accordingly, we take $a,b,c,d,e,a',b',c',d'>0$ and $\phi\in[0,2\pi)$.
The quark masses are
\begin{eqnarray}
\nonumber &&m_u \simeq \dfrac{abe}{cd} \varepsilon^7 v_u \,,~~~ m_c \simeq \dfrac{cd}{e}\varepsilon^4 v_u \,,~~~ m_t \simeq e v_u \,,  \\
&&m_d \simeq \dfrac{a' b'}{c'} \varepsilon^7 v_d \,,~~~ m_s \simeq c' \varepsilon^5 v_d \,,~~~ m_b \simeq d' \varepsilon^2 v_d \,.
\end{eqnarray}
The mixing angles and the CP violating phases are
\begin{eqnarray}
|V_{us}| \simeq \left| \dfrac{ae}{cd} + \dfrac{a'}{c'} e^{-i\phi} \right| \varepsilon \,,~~~ |V_{ub}| \simeq \dfrac{a}{d}\varepsilon^3 \,,~~~ |V_{cb}| \simeq \dfrac{c}{e} \varepsilon^2 \,,~~~\mathcal{J}_{\rm CP} \simeq - \dfrac{ac}{de} \dfrac{a'}{c'} \sin \phi\, \varepsilon^6 \,,
\end{eqnarray}
which gives
\begin{eqnarray}
\cot\delta^q_{CP}  \simeq - \cot \phi - \dfrac{a e}{c d} \dfrac{c'}{a'} \dfrac{1}{\sin \phi} \,.
\end{eqnarray}
In order to explain Yukawa hierarchies solely by non-invertible symmetry, the parameters $a \sim e$ and $a' \sim d'$ should be $\mathcal{O}(1)$. Here we choose to quantify it by introducing a measure $\chi^2_{\mathcal{O}(1)}$ defined as~\cite{Linster:2018avp}
\begin{eqnarray}\label{eq:chi_O1}
\chi^2_{\mathcal{O}(1)} = \sum_{i,j,f} \left( \log |c^{ij}_f| \right)^2
\end{eqnarray}
where $c_f^{ij}$ spans all the coefficients in the Yukawa matrices.
The function $\chi^2_{\mathcal{O}(1)}$ can be minimized at $\varepsilon \simeq 0.2141$. The best fit values of the parameters are
\begin{eqnarray}
\nonumber a &=& 0.8587 \,,~~~ b = 1.0332\,,~~~ c = 1.1434\,,~~~ d = 2.2239 \,,~~~ e = 1.2130\,,  \\
a' &=& 0.8247\,,~~~ b' = 0.9455\,,~~~ c' = 0.7990\,,~~~ d' = 0.4321\,,~~~\phi = 1.5095\pi
\end{eqnarray}
The predictions for quark masses and mixing parameters are given by
\begin{eqnarray}
\nonumber m_u &=& 0.001069\,{\rm GeV}\,,~~~ m_c = 0.5418\,{\rm GeV}\,,~~~ m_t = 149.9\,{\rm GeV}\,,~~~ \\
\nonumber m_d &=& 0.002350\,{\rm GeV}\,,~~~ m_s = 0.04660\,{\rm GeV}\,,~~~ m_b = 2.437\,{\rm GeV}\,,~~~ \\
\theta^q_{12} &=& 0.2270 \,,~~~ \theta^q_{13} = 0.003773 \,,~~~ \theta^q_{23} = 0.04271\,,~~~\delta^q_{CP} = 65.26^\circ \,.
\end{eqnarray}
As shown in table~\ref{tab:quark_data}, the predicted quark masses and mixing parameters are in excellent agreement with the experimental values.
It has recently been noted that the CP-violating phase $\phi$ in this structure is close to $3\pi/2$, a feature that is closely related to the unitary Yukawa triangle~\cite{Arkani-Hamed:2026uos}. In fact, the structure of the unitary triangle is directly reflected in simple properties of the Yukawa matrices. 
Moreover, this texture satisfies $\arg\det(Y_uY_d)=0$, thereby eliminating the Standard Model contribution to the physical QCD $\theta$ angle and providing a potential solution to the strong CP problem without axions.

For the lepton sector, we can also parameterize the mass matrices as follows,
\begin{eqnarray}
\nonumber M_e &=& \begin{pmatrix}
0 ~&~ a_e \varepsilon^6 ~&~ 0   \\
b_e \varepsilon^7 ~&~ c_e e^{i\varphi}\, \varepsilon^5 ~&~ 0   \\
0 ~&~ 0 ~&~ d_e \varepsilon^3
\end{pmatrix} v_d \,,~~~
M_D = \begin{pmatrix}
a_\nu \varepsilon^3 ~&~ 0   \\
0 ~&~ b_\nu \varepsilon^2 \\
c_\nu e^{i\rho} \varepsilon^2 ~&~ d_\nu \varepsilon^2
\end{pmatrix} v_u \,,~~~
M_N =\begin{pmatrix}
M_1 ~&~ 0 \\
0 ~&~ M_2 
\end{pmatrix} \\
M_\nu &=& \begin{pmatrix}
\dfrac{a_\nu^2}{M_1} \varepsilon^6 ~&~ 0 ~&~ \dfrac{a_\nu c_\nu e^{i\rho}}{M_1} \varepsilon^5 \\
0 ~&~ \dfrac{b_\nu^2}{M_2} \varepsilon^4 ~&~ \dfrac{b_\nu d_\nu}{M_2} \varepsilon^4 \\
\dfrac{a_\nu c_\nu e^{i\rho}}{M_1} \varepsilon^5 ~&~ \dfrac{b_\nu d_\nu}{M_2} \varepsilon^4 ~&~ \left(\dfrac{d_\nu^2}{M_2} + \dfrac{c_\nu^2 e^{2i\rho}}{M_1} \right) \varepsilon^4
\end{pmatrix} v_u^2\,.\label{eq:lepton-obs-U1-case3}
\end{eqnarray}
The lepton masses are
\begin{eqnarray}
\nonumber m_e &\simeq & \dfrac{a_e b_e}{c_e} \varepsilon^8 v_d\,,~~~ m_\mu \simeq c_e \varepsilon^5 v_d\,,~~~ m_\tau = d_e \varepsilon^3 v_d \,, \\  
m_1 &=& 0\,,~~~ m_{2,3} \simeq \sqrt{\dfrac{1}{2} \left( x \mp \sqrt{x^2 - 4y} \right)} \dfrac{\varepsilon^4 v_u^2}{M} \,.
\end{eqnarray}
where
\begin{eqnarray}
x = b_\nu^4 + 2 b_\nu^2 d_\nu^2 + | d_\nu^2 + c_\nu^2 \,e^{2i\rho} |^2 + 2 a_\nu^2 c_\nu^2 \varepsilon^2 \,,~~~
y = \left( b_\nu^2 c_\nu^2 + a_\nu^2 (b_\nu^2 + d_\nu^2) \varepsilon^2  \right)^2 \,,
\end{eqnarray}
and we set $M_1 = M_2 = M$ for simplicity~\footnote{The light neutrino mass matrix $M_{\nu}$ in Eq.~\eqref{eq:lepton-obs-U1-case3} only depend on the combinations $\frac{a^2_{\nu}}{M_1}$, $\frac{c^2_{\nu}}{M_1}$, $\frac{a_{\nu}c_{\nu}}{M_1}$, $\frac{b^2_{\nu}}{M_2}$, $\frac{d^2_{\nu}}{M_2}$ and $\frac{b_{\nu}d_{\nu}}{M_2}$ so that the values of heavy right-handed neutrino masses $M_1$ and $M_2$ can not be fixed by light neutrino masses and mixing angles. By rescaling $a_{\nu}$, $b_{\nu}$, $c_{\nu}$ and $d_{\nu}$, we can always choose $M_1=M_2$. We expect that their values could be determined when considering leptogenesis.}.
The mixing parameters are determined to be
\begin{eqnarray}
\nonumber \theta^{\ell}_{23} &=& \dfrac{1}{2} \arctan \dfrac{2b_\nu d_\nu |b_\nu^2 + d_\nu^2 + c_\nu^2 \, e^{2i\rho}|}{|d_\nu^2 + c_\nu^2 \, e^{2i\rho}|^2 - b_\nu^4} \,,  \\
\nonumber \sin^2 \theta^{\ell}_{13} &=& \left| \dfrac{a_e}{c_e} \sin \theta^{\ell}_{23} - \dfrac{a_\nu c_\nu \varepsilon^4 v_u^2}{m_3 M} \cos \theta^{\ell}_{23} \, e^{i(\varphi + \rho - \eta + 2\xi_3)} \right|^2  \varepsilon^2  \,,   \\
\nonumber \sin^2 \theta^{\ell}_{12} &=& \left| \dfrac{a_e \varepsilon}{c_e} \cos\theta^{\ell}_{23} \cos\psi + \sin \psi \, e^{i(\varphi + \rho - \eta + 2\xi_2)} \right|^2  \,,  \\
\nonumber \delta^\ell_{\rm CP} &=& \xi_3 - \xi_2 + \arg \left( \dfrac{a_e \varepsilon \cos\theta^\ell_{23}\cos\psi \, e^{i(\xi_2 - \varphi)} + c_e \sin\psi \, e^{i(\rho + \eta - \xi_2)} }{a_e \sin\theta^\ell_{23}\, e^{i(\xi_3-\varphi)} - a_\nu c_\nu c_e \cos\theta^\ell_{23} \,e^{i(\rho + \eta - \xi_3)}\varepsilon^4 v_u^2 /(m_3 M) } \right) \,,  \\
\alpha_{21} &=& 2 ( \xi_2 - \xi_3) \,.\label{eq:lepton-pars-modelIII}
\end{eqnarray}
where 
\begin{eqnarray}
\eta &=& \arg \left( b_\nu^2 + d_\nu^2 + c_\nu^2 e^{-2i\rho} \right) \,,~~~
\psi = \arctan \dfrac{a_\nu \varepsilon}{c_\nu \sin\theta^\ell_{23}} \,, \\
\xi_2 &=& \dfrac{1}{2} \arg \left( c_\nu^2 \sin^2 \theta^\nu_{23} \, e^{2i(\eta + \rho)} + \left( b_\nu \cos \theta^\nu_{23} - d_\nu \sin \theta^\nu_{23} \, e^{i\eta} \right)^2 \right)   \\
\xi_3 &=& \dfrac{1}{2} \arg \left( c_\nu^2 \cos^2 \theta^\nu_{23} \, e^{2i(\eta + \rho)} + \left( b_\nu \sin \theta^\nu_{23} + d_\nu \cos \theta^\nu_{23} \, e^{i\eta} \right)^2 \right) \,.
\end{eqnarray}
Obtaining a large solar mixing angle $\sin^2\theta_{12}$ requires an enhancement from the coefficient ratios, while reproducing the neutrino mass ratio $m_2/m_3\sim\varepsilon$ requires a suitable tuning of the coefficients. These two requirements imply that the combination $\dfrac{a_\nu \varepsilon}{c_\nu \sin \theta_{23}}$ should be of order $1$.
Assuming the sequential dominance of right handed neutrinos, $a^2_{\nu}\varepsilon^2/M_1, c^2_{\nu}/M_1, a_{\nu}c_{\nu}\varepsilon/M_1\ll b^2_{\nu}/M_2, d^2_{\nu}/M_2, b_{\nu}d_{\nu}/M_2$, the above formulae have a simple approximate form,
\begin{eqnarray}
\nonumber m_1 &=& 0 \,,~~~ m_2 \approx \left( a_\nu^2\varepsilon^2 + c_\nu^2 \sin^2 \theta^\ell_{23}  \right) \dfrac{\varepsilon^4 v_u^2}{M_1} \,,~~~ m_3 \approx \left( b_\nu^2 + d_\nu^2 \right) \dfrac{\varepsilon^4 v_u^2}{M_2} \,,  \\
\tan \theta^\ell_{23} &\approx & \dfrac{b_\nu}{d_\nu}\,,~~~ \tan \theta^\ell_{12} \approx  \dfrac{a_\nu \varepsilon}{c_\nu \sin\theta^\ell_{23}} \,,~~~ \theta^\ell_{13} \approx \left| \dfrac{a_e}{c_e} \sin\theta^\ell_{23} + \dfrac{\varepsilon^4 v_u^2}{m_3} \dfrac{a_\nu c_\nu}{M_1} \cos\theta^\ell_{23} \,e^{i(\rho+\varphi)}  \right| \varepsilon \,. ~~\label{eq:SD-limit}
\end{eqnarray}
Ignoring charged lepton corrections, the above approximate formulae are consistent with the general results of sequential dominance given in~\cite{King:2002nf}. The best fit values of the couplings with $\varepsilon \simeq 0.2141$ are determined to be~\footnote{Notice the sequential dominance limit is not well satisfied at the best fit point, and the approximation formulas of Eq.~\eqref{eq:SD-limit} are accurate up to twenty percent.} 
\begin{eqnarray}
\nonumber a_e &=& 1.2934\,,~~~b_e = 1.3054\,,~~~ c_e = 1.8465\,,~~~ d_e = 1.4549\,,~~~\varphi = 1.2313 \pi \,, \\
 a_\nu &=& 1.5939\,,~~~ b_\nu = 0.881854\,,~~~ c_\nu = 0.567501\,,~~~ d_\nu = 0.874065\,,~~~ \rho = 0.25409 \pi \,,~~~~
\end{eqnarray}
where we take $M_1 = M_2 =M =10^{12}\,{\rm GeV}$.The observables are
\begin{eqnarray}
\nonumber \sin^2 \theta^{\ell}_{12} &=& 0.3088 \,,~~~ \sin^2 \theta^{\ell}_{23} = 0.4698\,,~~~ \sin^2 \theta^{\ell}_{13} = 0.02248\,,~~~\delta^\ell_{\rm CP} = 1.189 \pi \,,  \\
\nonumber ~~~ \alpha_{21} &=& 1.549 \pi \,,~~~ m_1 = 0  \,,~~~ m_2 = 8.681 \,{\rm meV}\,,~~~ m_3 = 50.110\,{\rm meV}\,, \\
\nonumber m_e &=& 0.4910 \,{\rm MeV} \,,~~~ m_\mu = 103.4\,{\rm MeV}\,,~~~ m_\tau = 1756.8\,{\rm MeV}\,, \\
m_\beta &=& 8.899 \,{\rm meV} \,,~~~ m_{\beta\beta} = 3.726\,{\rm meV} \,.
\end{eqnarray}

Here, the effective Majorana neutrino mass $m_{\beta\beta}$ is defined as
\begin{eqnarray}
m_{\beta\beta} = \left| m_2 \sin^2 \theta^{\ell}_{12} \cos^2 \theta^{\ell}_{13} \, e^{i\alpha_{21}} + m_3 \sin^2 \theta^{\ell}_{13} \, e^{-2i \delta^\ell_{\rm CP}} \right| \,,~~~ m_1 = 0 \,.
\end{eqnarray}
The kinematical mass $m_\beta$ is defined as
\begin{eqnarray}
m_\beta = \left( m_2^2 \sin^2 \theta^{\ell}_{12} \cos^2 \theta^{\ell}_{13} + m_3^2 \sin^2 \theta^{\ell}_{13} \right)^{1/2} \,.
\end{eqnarray}
For neutrinoless double-beta decay, KamLAND-Zen currently provides the most stringent constraint, $m_{\beta\beta}<(28-122)\,{\rm meV}$ at $90\%$ C.L.~\cite{KamLAND-Zen:2024eml}. Future tonne-scale experiments, including LEGEND-1000~\cite{LEGEND:2021bnm} and nEXO~\cite{nEXO:2021ujk}, are expected to reach sensitivities of $m_{\beta\beta}\sim(9-21)\,{\rm meV}$ and $(4.7-20.3)\,{\rm meV}$, respectively, for a ten-year exposure. The current KATRIN bound on the effective kinematic mass is $m_\beta<0.45\,{\rm eV}$ at $90\%$ C.L.~\cite{KATRIN:2024cdt}, with projected sensitivities of approximately $0.2\,{\rm eV}$ for KATRIN~\cite{KATRIN:2021dfa} and $0.04\,{\rm eV}$ for Project 8~\cite{Project8:2022wqh}. Moreover the model predicts neutrino mass sum $\sum_i m_i \simeq 58.791\,{\rm meV}$ which satisfies the current cosmological bound $\sum_i m_i<120\,{\rm meV}$ from Planck~\cite{Planck:2018vyg}. Interestingly, it falls within the projected sensitivity of future cosmological observations combining Euclid+CMBS4+LiteBIRD, which are expected to probe $\sum_i m_i$ at the level of $(44-76)\,{\rm meV}$~\cite{Euclid:2024imf}. Therefore the predicted values of $m_\beta$ and $m_{\beta\beta}$ in these models remain below the sensitivity of current and near-future laboratory experiments, while the predicted neutrino mass sum may be probed by upcoming cosmological observations.

\begin{figure}[hptb!]
\centering
\includegraphics[width=6.5in]{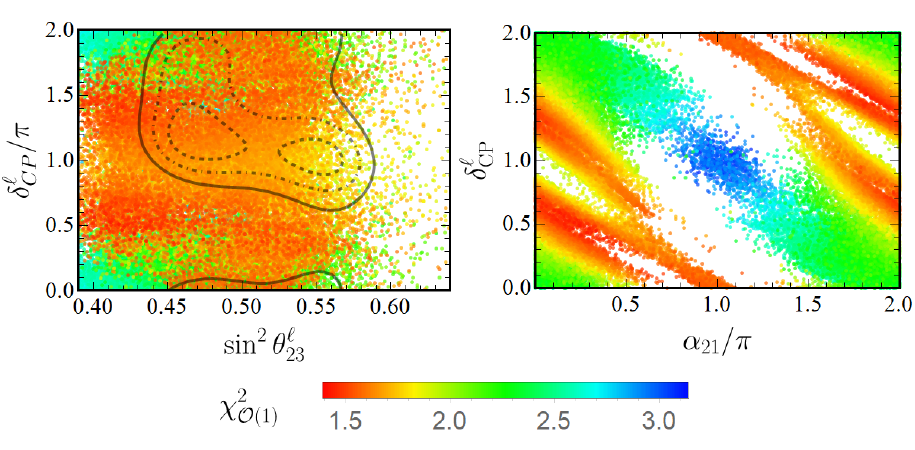}
\caption{\label{fig:case3_corr}Predictions for correlations between the mixing angles and CP violation phases for the Model III. The black dashed line, dash-dotted line and solid line denote the $1\sigma$, $2\sigma$ and $3\sigma$ contours respectively in the $\sin^2\theta^{\ell}_{23}-\delta^{\ell}_{CP}$ plane~\cite{Esteban:2024eli}.}
\end{figure}

Furthermore, we employ the sampler \texttt{MultiNest}~\cite{Feroz:2007kg,Feroz:2008xx} to scan the parameter space, imposing that all measured lepton observables lie within their experimentally preferred $3\sigma$ ranges. We find that the predicted allowed interval of $m_{\beta\beta}$ is $[1.231\,{\rm meV}, 4.013\,{\rm meV}]$ and $m_\beta$ is $[8.389\,{\rm meV}, 9.407\,{\rm meV}]$. 
Analogously to the quark sector, we can use Eq.~\eqref{eq:chi_O1} to define the $\chi_{\mathcal{O}(1)}^2$ function for the lepton sector, where only the Yukawa couplings in the charged-lepton and neutrino Dirac mass matrices need to be taken into account. The $\chi_{\mathcal{O}(1)}^2$ function can be used as a measure of the naturalness of the model, since all the couplings are expected to be order one. For this model, $\chi_{\mathcal{O}(1)}^2$ has a lower bound of approximately $1.2$.
As can be seen from figure~\ref{fig:case3_corr} the atmospheric mixing angle $\theta^\ell_{23}$ tends to lie in the first octant. This is because, for the $\mathcal{O}(1)$ parameters, reproducing the light-neutrino mass hierarchy together with a large solar mixing angle $\theta^\ell_{12}$ requires a slightly smaller atmospheric mixing angle, as can be seen from Eq.~\eqref{eq:lepton-pars-modelIII}. The Dirac CP-violating phase $\delta^\ell_{\rm CP}$ and the Majorana phase $\alpha_{21}$ exhibit a strong correlation.

\section{Conclusion\label{sec:conclusion}}

Non-invertible symmetries offer new possibilities for addressing the flavor puzzle by extending the selection rules available in conventional group-based constructions. Their fusion rules can restrict Yukawa interactions and Majorana mass terms, providing a symmetry origin for texture zeros in the quark and lepton sectors~\cite{Kobayashi:2024cvp,Kobayashi:2025ldi,Jiang:2025psz}. Such restrictions can lead to relations among fermion masses, mixing angles and CP-violating phases, while suitable quark textures have also motivated possible axion-less approaches to the strong CP problem~\cite{Liang:2025dkm}. Beyond determining which couplings vanish, an important question is whether these symmetries can explain the hierarchical sizes of the nonzero couplings. This motivates their incorporation into the Froggatt--Nielsen mechanism, linking texture selection to the origin of fermion mass hierarchies. In this work, we have formulated the FN mechanism for non-invertible symmetries obtained by $\mathbb{Z}_2$ gauging of $\mathbb{Z}_N$ and $U(1)$. Here, $\mathbb{Z}_2$ gauging of $U(1)$ refers to gauging the inversion automorphism $g\mapsto g^{-1}$, which groups opposite labels into the same class $[k]=\{e^{ik\theta},e^{-ik\theta}\}$. The resulting fusion rules admit both sum and difference channels. One of the central results of section~\ref{sec:framework-FN} is an analytic prescription for determining the leading power of $\varepsilon=\langle S\rangle/\Lambda$. Specifically, the exponent is fixed by the minimum number of flavon insertions necessary for the corresponding interaction to contain the identity fusion channel. For a unit flavon assignment and ordered fermion and Higgs labels $k_1\leq k_2\leq k_3$, this prescription yields $n_{ij}=|k_3-k_2-k_1|$ for the Yukawa exponent in the $U(1)$ construction. For more general flavon assignments, additional divisibility conditions must be satisfied; if no number of flavon insertions satisfies these conditions, the corresponding Yukawa entry is forbidden. Consequently, both the hierarchical suppression factors and the texture-zero structure are fixed analytically before performing any numerical fit.
In conventional FN models, the suppression power follows from compensating the fixed total Abelian charge of an operator. Here it is determined by the least suppressed allowed fusion channel, which can involve different relative signs of the field labels. Nevertheless, for a trivial Higgs assignment and a unit flavon label, the suppression powers of the Yukawa couplings coincide with those of conventional Abelian models. The distinction is particularly clear for Majorana terms. With a unit flavon label, the non-invertible rule gives $n^N_{ij}=|k_{\nu^c_{Ri}}-k_{\nu^c_{Rj}}|$, so diagonal masses are unsuppressed even for nonzero labels, whereas conventional FN charge sums can suppress them. 

In order to illustrate how this framework accommodates fermion mass hierarchies and small quark mixing angles, we have constructed one model based on $Z_2$ gauging of $Z_{12}$ and three benchmark models based on $Z_2$ gauging of $U(1)$ symmetry. In particular, the $\mathbb{Z}_2$ gauging of $\mathbb{Z}_{12}$ model realizes hierarchical quark and lepton mass matrices without texture zeros, with FN suppressions accounting for the charged fermion mass hierarchies and the mixing angles. The benchmark Model I based on $\mathbb{Z}_2$ gauging of $U(1)$ has hierarchical Yukawa matrices without texture zeros, whereas then benchmark Model II combines hierarchical suppressions with symmetry-enforced zeros. Random scans of $\mathcal{O}(1)$ coefficients support the ability of these representative textures to reproduce the observed quark hierarchies at the order-of-magnitude level. The benchmark Model III demonstrates the combined predictive role of non-invertible symmetry and the FN mechanism particularly clearly. The non-invertible selection rules enforce nine texture zeros in the quark sector, while the FN mechanism fixes the leading power of $\varepsilon$ in every allowed mass-matrix entry. Thus, the construction determines both the locations of the zeros and the order of magnitude of each nonzero element, up to $\mathcal{O}(1)$ coefficients and the relevant overall mass scales. The same principle governs the charged lepton, Dirac neutrino and right-handed Majorana mass matrices. This provides a symmetry-based explanation of the hierarchical entries in addition to the texture-zero pattern itself. Allowing independent, non-Hermitian Yukawa entries yields an excellent fit to the quark masses and CKM parameters with $\mathcal{O}(1)$ coefficients and $\varepsilon\simeq0.214$, including the pattern $|V_{us}|\sim\varepsilon$, $|V_{cb}|\sim\varepsilon^2$ and $|V_{ub}|\sim\varepsilon^3$. In the displayed parametrization of $M_{u,d}$ in Eq~\eqref{eq:ModelIII_Mud}, the single CP-violating phase $\phi$ is close to $3\pi/2$, while $\arg\det(Y_uY_d)=0$. This structure motivates further investigation of a possible connection to the strong CP problem, including its ultraviolet origin and radiative stability. In the lepton sector, the type-I seesaw mechanism generates small neutrino masses together with large atmospheric mixing and a suppressed reactor angle. Reproducing the solar mixing angle and the neutrino mass ratio nevertheless requires suitable coefficient ratios. The minimal seesaw realization in Model III has normal ordering with $m_1=0$. These benchmark values illustrate the phenomenology of the texture rather than parameter-independent predictions.

A further direction is to test the FN origin of flavor through processes beyond fermion masses and mixing. Conventional FN studies show that viable quark charge assignments can produce distinct patterns of flavor-changing four-quark operators, making correlations among $\Delta F=2$ meson-mixing observables sensitive probes of the underlying flavor structure~\cite{Cornella:2023zme}. In the lepton sector, correlated signals in $\mu\to e\gamma$, $\mu\to 3e$ and $\mu$--$e$ conversion in nuclei can provide complementary discrimination among FN textures, with neutrinoless double-beta decay further probing Majorana scenarios~\cite{Cornella:2024jaw}. These results motivate extending our analysis to higher-dimensional flavor-violating operators. In the present framework, their allowed structures and leading FN powers must be derived from the non-invertible fusion rules and then rotated to the fermion mass basis; Yukawa texture zeros alone do not imply the absence of the corresponding flavor-changing amplitudes. A joint analysis of quark, lepton and semileptonic operators could therefore test whether the additional selection rules yield correlations that distinguish these constructions from conventional FN models and from one another. Quantitative rates require specifying the relevant heavy scales and Wilson coefficients, including their dependence on the ultraviolet completion, and evolving the operators to the experimental scales. Such an analysis would turn the successful mass textures into a broader program of flavor tests.

The flavon sector also connects the origin of fermion mass hierarchies to the evolution of the early Universe. As discussed in appendix~\ref{app:VEV-H-S}, its coupling to the Higgs links the flavor-breaking sector to the thermal scalar potential. Thermal corrections may restore the non-invertible symmetry at high temperatures, followed by spontaneous breaking as the Universe cools. The resulting evolution of the flavon expectation value would make the FN suppression factors temperature dependent, connecting the emergence of flavor hierarchies to the cosmological symmetry-breaking history. If the transition is first order and completes, bubble nucleation and the subsequent dynamics may generate a stochastic gravitational-wave background, offering a complementary probe of the flavor-breaking scale and scalar interactions. These possibilities are conditional: the zero-temperature vacuum analysis alone establishes neither symmetry restoration nor the order or completion of the transition. A finite-temperature study of the coupled Higgs--flavon potential, followed by an analysis of transition dynamics and the resulting gravitational-wave spectrum, is therefore an important next step. Since $S$ acquires a nonzero VEV and its physical excitation can mix with the Higgs and decay, it does not provide a stable dark matter candidate in the present construction. Its cosmological significance instead lies in how flavor symmetry breaking may shape the early thermal history.

In summary, we have reformulated the FN mechanism within the framework of non-invertible symmetries obtained by $\mathbb{Z}_2$ gauging of $\mathbb{Z}_N$ and $U(1)$. We have derived a master formula for the leading power of flavon insertions, providing a systematic prescription for determining both the hierarchical of allowed interactions and the texture zeros enforced by the fusion rules. In particular, we have constructed a predictive realization, the benchmark Model III based on $\mathbb{Z}_2$ gauging of $U(1)$ symmetry, with nine texture zeros across the up and down quark Yukawa matrices. Beyond fixing the zero pattern, the FN mechanism determines the parametric size of every nonzero entry up to $\mathcal{O}(1)$ coefficients, allowing a successful description of quark masses and CKM mixing. Using the minimal seesaw mechanism, the model can accommodate the observed lepton masses and mixing, and the neutrino mass spectrum is predicted to be normal ordering with one massless state. These results demonstrate how non-invertible selection rules and FN suppressions can jointly account for the structure and hierarchies of fermion mass matrices, providing a basis for further flavor and cosmological tests.

\vskip0.2in 
{\bf Note Added:} At the very final stages of the completion of this work a paper~\cite{Otsuka:2026fpy} has been submitted to the arXiv that explores the connection between non-invertible symmetries and the Froggatt--Nielsen mechanism. Their construction uses a non-invertible completion of an anomalous FN symmetry and hidden-sector instanton effects to generate suppressed Dirac-neutrino Yukawa couplings. The present work develops the FN mechanism using fusion rules arising from $\mathbb{Z}_2$ gauging of $\mathbb{Z}_N$ and $U(1)$, deriving the leading flavon powers and symmetry-enforced texture zeros. The two studies thus address a related theme through different symmetry constructions and neutrino-mass mechanisms.

\section*{Acknowledgements}

BYQ and GJD are supported by the National Natural Science Foundation of China under Grant Nos. 12375104, 12547106 and Guizhou Provincial Major Scientific and Technological Program XKBF (2025)010. SFK acknowledges the STFC Consolidated Grant ST/X000583/1 and his work was funded by a Leverhulme Trust Emeritus Fellowship Grant.

\newpage

\begin{appendix}

\section{\label{app:VEV-H-S} Vacuum structure of the scalar potential compatible with $\mathbb{Z}_2$ gauging of $\mathbb{Z}_N$ (or $U(1)$) symmetry}

In this work, the $\mathbb{Z}_2$ gauging of the $\mathbb{Z}_N$ symmetry is spontaneously broken by the vacuum expectation value of a real scalar field $S$, which is a SM singlet assigned to the equivalence class $[k_S]$. The SM Higgs field $H$ is assigned to the equivalence class $[k_H]$. The scalar potential involving $H$ and $S$, invariant under the $\mathbb{Z}_2$ gauging of $\mathbb{Z}_N$ symmetry, must contain at least the following terms:
\begin{eqnarray}\label{eq:potential}
 V(H, S) &=& - \mu_H^2 H^\dag H + \lambda_H (H^\dag H)^2 - \mu_S^2 S^2 + \lambda_S S^4 + \lambda (H^\dag H) S^2\,.
\end{eqnarray}
This potential has been discussed in the literature such as Refs.~\cite{Burgess:2000yq,Baldes:2016gaf}.
Additional terms in the potential may arise if the assignments  $[k_H]$ and $[k_S]$ satisfy some certain relations, as given below 
\begin{eqnarray}
\nonumber [2k_{H}] = [k_{S}] &:&~~ \mu_{HS} H^\dag H S + \text{h.c.}  \\
~[3k_{S}]= [0] &:&~~ \mu_S SSS + \text{h.c.}
\end{eqnarray}
The above results for the scalar potential are unchanged when the $U(1)$ symmetry, rather than the $\mathbb{Z}_N$  symmetry, is gauged by $\mathbb{Z}_2$.

In order to prevent the potential acquiring large negative values at large field values thus the scalar potential $V(H, S)$ is bounded from below, the following conditions should be obeyed by the quartic couplings of this model~\cite{Kannike:2012pe}, 
\begin{eqnarray}
\label{eq:copositive-cond}\lambda_H > 0\,,~~~ \lambda_S > 0\,,~~~ \lambda > - 2 \sqrt{\lambda_H \lambda_S}\,.
\end{eqnarray}
Let $v_H$ and $v_S$ be the vacuum expectation values of $H$ and $S$, respectively, at the global minimum of the scalar potential. In the unitary gauge, $H$ and $S$ can be parametrized as
\begin{eqnarray}
H =\dfrac{1}{\sqrt{2}}\begin{pmatrix}
0  \\
\phi 
\end{pmatrix}
= \dfrac{1}{\sqrt{2}}\begin{pmatrix}
0  \\
v_H + h 
\end{pmatrix} \,,~~~
S = \dfrac{1}{\sqrt{2}} \sigma = \dfrac{1}{\sqrt{2}} ( v_S + s )\,.
\end{eqnarray}
Then the scalar potential in Eq.~\eqref{eq:potential} can be written as
\begin{eqnarray}
V(\phi, \sigma) = - \dfrac{\mu_H^2}{2} \phi^2  + \dfrac{\lambda_H}{4} \phi^4 - \dfrac{\mu_S^2}{2}  \sigma^2 + \dfrac{\lambda_S}{4} \sigma^4 + \dfrac{\lambda}{4} \phi^2 \sigma^2\,.
\end{eqnarray}
The minimization conditions of the scalar potential are given by
\begin{eqnarray}
\dfrac{\partial V}{\partial \phi} = \lambda_H v_H^3 + \dfrac{\lambda}{2} v_S^2 v_H - \mu_H^2 v_H &=& 0 \,, \\
\dfrac{\partial V}{\partial \sigma} = \lambda_S v_S^3 + \dfrac{\lambda}{2} v_H^2 v_S - \mu_S^2 v_S &=& 0 \,,
\end{eqnarray}
which implies
\begin{eqnarray}
\label{eq:vH-vS}v_H^2 = \dfrac{4\lambda_S \mu_H^2 - 2\lambda \mu_S^2}{4\lambda_H\lambda_S - \lambda^2}  \,,~~~
v_S^2 = \dfrac{4\lambda_H \mu_S^2 - 2\lambda \mu_H^2}{4\lambda_H\lambda_S - \lambda^2} \,.
\end{eqnarray}
The squared VEVs $v^2_H$ and $v^2_S$ must be positive, which requires
\begin{eqnarray}
\label{eq:VEV_existence} && ( 4\lambda_S \mu_H^2 - 2\lambda \mu_S^2 ) (4\lambda_H\lambda_S - \lambda^2) > 0 \,, \quad ~~ ( 4\lambda_H \mu_S^2 - 2\lambda \mu_H^2 ) (4\lambda_H\lambda_S - \lambda^2) > 0\,.
\end{eqnarray}
Combining the positivity conditions for $v_H^2$ and $v_S^2$ with Eq.~\eqref{eq:copositive-cond}, we obtain the following constraints on the parameters. For $\mu_H^2>0$ and $\mu_S^2<0$,
\begin{eqnarray}
\label{eq:VEV-cond1}
\frac{2\lambda_S\mu_H^2}{\mu_S^2}
<-2\sqrt{\lambda_H\lambda_S}
<\lambda<
\frac{2\lambda_H\mu_S^2}{\mu_H^2}\,,
\end{eqnarray}
while for $\mu_H^2<0$ and $\mu_S^2>0$,
\begin{eqnarray}
\label{eq:VEV-cond2}\frac{2\lambda_H\mu_S^2}{\mu_H^2}
<-2\sqrt{\lambda_H\lambda_S}
<\lambda<
\frac{2\lambda_S\mu_H^2}{\mu_S^2}\,.
\end{eqnarray}
For $\mu_H^2>0$ and $\mu_S^2>0$, we instead have
\begin{eqnarray}
\label{eq:VEV-cond3}
-2\sqrt{\lambda_H\lambda_S}<\lambda<
\min\left\{
\frac{2\lambda_S\mu_H^2}{\mu_S^2},
\frac{2\lambda_H\mu_S^2}{\mu_H^2}
\right\}
\quad\text{or}\quad
\lambda>
\max\left\{
\frac{2\lambda_S\mu_H^2}{\mu_S^2},
\frac{2\lambda_H\mu_S^2}{\mu_H^2}
\right\}\,.
\end{eqnarray}
For $\mu_H^2<0$ and $\mu_S^2<0$, however, non-zero VEVs are incompatible with Eq.~\eqref{eq:copositive-cond}. 

The solution in Eq.~\eqref{eq:vH-vS} represents a stable vacuum only if the scalar potential has a local minimum at this point. Hence, the Hessian matrix $\mathcal{H}$ evaluated at the vacuum must be positive definite. It takes the following form,
\begin{eqnarray}
\mathcal{H}\equiv\begin{pmatrix}
\dfrac{\partial^2 V}{\partial \phi^2}  ~&~ \dfrac{\partial^2 V}{\partial \phi\partial \sigma} \\
\dfrac{\partial^2 V}{\partial \phi\partial \sigma}  ~&~ \dfrac{\partial^2 V}{\partial \sigma^2}
\end{pmatrix}=\begin{pmatrix}
3\lambda_H v_H^2 + \dfrac{\lambda}{2} v_S^2 - \mu_H^2 ~&~ \lambda v_H v_S  \\
\lambda v_H v_S ~&~ 3 \lambda_S v_S^2 + \dfrac{\lambda}{2} v_H^2 - \mu_S^2
\end{pmatrix}\,,
\end{eqnarray}
which is a $2\times2$ real symmetric matrix. Its mass eigenvalues are determined to be 
\begin{eqnarray}
m^2_{\pm} = \dfrac{1}{2} \left(\mathcal{H}_{11} +\mathcal{H}_{22}\pm \sqrt{(\mathcal{H}_{11} - \mathcal{H}_{22})^2 + 4 \mathcal{H}_{12}^2} \right)\,,
\end{eqnarray}
where $\mathcal{H}_{ij}$ refers to the $(ij)$ entry of $\mathcal{H}$. The stability of the vacuum requires both eigenvalues of the Hessian matrix are positive, i.e. $m_{\pm}^2 > 0$. Then we have
\begin{eqnarray}
\dfrac{2\lambda_H\lambda_S(\mu_H^2 + \mu_S^2) -\lambda(\lambda_S \mu_H^2 + \lambda_H \mu_S^2) }{4\lambda_H\lambda_S - \lambda^2} > 0 \,,~\quad 
\dfrac{(2\lambda_H \mu_S^2 - \lambda \mu_H^2)(2\lambda_S \mu_H^2 - \lambda \mu_S^2)}{4\lambda_H \lambda_S - \lambda^2}  > 0\,.\quad
\end{eqnarray} 
After imposing the above inequalities together with the conditions in Eqs.~\eqref{eq:VEV-cond1},\eqref{eq:VEV-cond2},\eqref{eq:VEV-cond3}, we find that a global minimum with $v_H\neq0$ and $v_S\neq0$  exists in the following parameter regions:
\begin{eqnarray}
\mu_H^2 > 0\,,~\mu_S^2 > 0 ~&:&~ -2\sqrt{\lambda_H\lambda_S} < \lambda < \min \left\{ \dfrac{2\lambda_S \mu_H^2}{\mu_S^2}, \dfrac{2\lambda_H \mu_S^2}{\mu_H^2} \right\}\,,\label{eq:lambda-cond1}   \\
\mu_H^2 > 0\,,~\mu_S^2 < 0 ~&:&~ -2\sqrt{\lambda_H\lambda_S} < \lambda < \dfrac{2\lambda_H \mu_S^2}{\mu_H^2}\,,  \label{eq:lambda-cond2}  \\
\mu_H^2 < 0\,,~\mu_S^2 > 0 ~&:&~ -2\sqrt{\lambda_H\lambda_S} < \lambda < \dfrac{2\lambda_S \mu_H^2}{\mu_S^2}\,. \label{eq:lambda-cond3} 
\end{eqnarray}
In the decoupling limit $\lambda\rightarrow0$, from Eq.~\eqref{eq:vH-vS} we have 
\begin{eqnarray}
v_H^2\approx\dfrac{\mu_H^2}{\lambda_H}  \,,~~~
v_S^2\approx\dfrac{\mu_S^2}{\lambda_S} \,,
\end{eqnarray}
The stability condition of \eqref{eq:lambda-cond1} is satisfied, and we have $v^2_S\gg v^2_H$ for $\mu^2_S\gg \mu^2_H$.  
\begin{eqnarray}
m^2_{+}\approx 2\mu_S^2 \,,~~~ m^2_{-}\approx 2 \mu_H^2 \,. 
\end{eqnarray}
It is worth emphasizing that, we don't get additional Goldstone boson besides the usual standard model Higgs mechanism.  Although the potential in Eq.~\eqref{eq:potential} contains only even powers of $S$, this does not make the flavon a stable dark matter candidate in the vacuum considered here. The FN mechanism requires $\langle S\rangle\neq0$, which spontaneously breaks the parity $S\to-S$ of this potential. With the field normalization used above, expanding the Higgs-portal term around the vacuum generates the interaction terms $\frac{1}{2}\lambda v_H\,hs^2 + \frac{1}{2}\lambda v_S\,h^2s$. Expanding the FN operators also generates couplings of the flavon excitation $s$ to fermions. Consequently, the physical scalar states can decay into kinematically accessible SM final states, and the non-invertible symmetry does not protect a stable flavon in this broken phase. The present construction therefore does not realize scalar singlet dark matter, and relic-density or dark-matter detection bounds derived for a stable singlet cannot be directly applied to it.

Moreover, the flavon field $S$ provides a direct link between flavor physics and the thermal history of the early Universe through its coupling to the Higgs field. At sufficiently high temperatures, thermal corrections to the scalar potential may restore the non-invertible symmetry, which is subsequently broken spontaneously as the Universe cools belows the critical temperature. This symmetry-breaking phase transition may generate a stochastic gravitational-wave background, thereby offering a complementary cosmological probe of the non-invertible symmetry and the underlying origin of flavor. Moreover, only even power os $S$ appears in the scalar potential in Eq.~\eqref{eq:potential}, consequently $S$ could be a dark matter candidate. A detailed analysis of the thermal history and the resulting gravitational-wave signal and dark matter signature is left for future work.

\section{\label{app:models-Z2-gauging-ZN14-ZN36} Other models based on $\mathbb{Z}_2$ gauging of $\mathbb{Z}_N$ symmetry }

In this appendix, we demonstrate that texture zeros in the Yukawa couplings can arise from the $\mathbb{Z}_2$ gauging of a $\mathbb{Z}_N$ symmetry. However, realizing the desired texture zeros typically requires a relatively large value of $N$. We therefore consider $N=36$, for which the corresponding equivalence classes are given by
\begin{eqnarray}
[0] = \{ e \}\,,~~~ [18] = \{ g^{18} \}\,,~~~  [k] = \{g^k, g^{-k} \} ~~~\text{for } 1\leq k \leq 17 \,.
\end{eqnarray}
The scalar field are assigned to be $H\sim [1]$ and $S\sim [3]$ respectively. The quark fields are assigned to be
\begin{eqnarray}
(Q_1, Q_2, Q_3) \sim ([0], [1], [9]) \,, (u_1, u_2, u_3) \sim ([17], [9], [8]) \,, (d_1, d_2, d_3) \sim ([17], [14], [14])\,.~~~~
\end{eqnarray}
With these assignment the mass matrices have the order of magnitude entries
\begin{eqnarray}
M_u \sim \begin{pmatrix}
\varepsilon^6 ~&~ 0 ~&~ \varepsilon^3   \\
\varepsilon^5 ~&~ \varepsilon^3 ~&~ \varepsilon^2   \\
\varepsilon^3 ~&~ 0 ~&~ 1
\end{pmatrix} v_H \,,~~~
M_d \sim \begin{pmatrix}
\varepsilon^6 ~&~ \varepsilon^5 ~&~ \varepsilon^5   \\
\varepsilon^5 ~&~ \varepsilon^4 ~&~ \varepsilon^4   \\
\varepsilon^3 ~&~ \varepsilon^2 ~&~ \varepsilon^2
\end{pmatrix} v_H \,.
\end{eqnarray}
In this case, the $(12)$ and $(32)$ entries of $M_{u}$ are vanishing, while the down type quark mass matrix $M_d$ has a same texture as that in Eq.~\eqref{eq:Mu-Md-N12}. The quark masses and CKM matrix elements can be estimated as follows, 
\begin{eqnarray}
\nonumber&& m_u\sim\varepsilon^{6}v_H\,,~~m_c\sim\varepsilon^{3}v_H\,,~~m_t\sim v_H\,,\\
\nonumber && m_d\sim \varepsilon^{6}v_H\,,~~m_s\sim \varepsilon^{4}v_H\,,~~m_b\sim \varepsilon^{2}v_H\,,\\
&&V_{us}\sim\varepsilon\,,~~V_{ub}\sim\varepsilon^3\,,~~V_{cb}\sim\varepsilon^2\,.
\end{eqnarray}
The Cabibbo mixing angle dominantly arises from the diagonalization of the $M_d$. Analogously the lepton fields are assigned to 
\begin{eqnarray}
\ell \sim ([0], [1], [9])\,,~~~ E_R \sim ([17], [14], [14])\,,~~~ \nu_R \sim ([3], [5], [7])
\end{eqnarray}
The charged lepton mass and neutrino mass matrices are
\begin{eqnarray}
\nonumber M_e &\sim & \begin{pmatrix}
\varepsilon^6 ~&~ \varepsilon^5 ~&~ \varepsilon^5   \\
\varepsilon^5 ~&~ \varepsilon^4 ~&~ \varepsilon^4   \\
\varepsilon^3 ~&~ \varepsilon^2 ~&~ \varepsilon^2
\end{pmatrix} v_H \,,~~~
M_D \sim  \begin{pmatrix}
0 ~&~ \varepsilon^2 ~&~ \varepsilon^2   \\
\varepsilon ~&~ \varepsilon ~&~ \varepsilon^3   \\
0 ~&~ \varepsilon ~&~ \varepsilon
\end{pmatrix} v_H \,,~~~   
M_N \sim \begin{pmatrix}
1 ~&~ 0 ~&~ 0   \\
0 ~&~ 1 ~&~ \varepsilon^4   \\
0 ~&~ \varepsilon^4 ~&~ 1
\end{pmatrix} M \,,~~~   \\
M_N^{-1} &\sim & \begin{pmatrix}
1 ~&~ 0 ~&~ 0   \\
0 ~&~ 1 ~&~ \varepsilon^4   \\
0 ~&~ \varepsilon^4 ~&~ 1
\end{pmatrix} M^{-1} \,,~~~
M_\nu \sim \begin{pmatrix}
\varepsilon^4 ~&~ \varepsilon^3 ~&~ \varepsilon^3   \\
\varepsilon^3 ~&~ \varepsilon^2 ~&~ \varepsilon^2   \\
\varepsilon^3 ~&~ \varepsilon^2 ~&~ \varepsilon^2
\end{pmatrix} \dfrac{v_H^2}{M} \,.
\end{eqnarray}
The order of magnitudes predictions for the charged lepton masses and neutrino mixing angels parameters,
\begin{eqnarray}
\nonumber&& m_e\sim\varepsilon^6 v_H\,,~~ m_\mu\sim\varepsilon^4 v_H\,,~~m_{\tau}\sim\varepsilon^2 v_H\,, \\
\nonumber &&\sin\theta_{12}\sim\varepsilon\,,~~\sin\theta_{13}\sim\varepsilon \,,~~\sin\theta_{23}\sim1 \,,  \\
&& m_1\sim \varepsilon^4\dfrac{v_H^2}{M}\,,~m_2 \sim \varepsilon^2\dfrac{v_H^2}{M}\,,~ m_3 \sim \varepsilon^2\dfrac{v_H^2}{M}\,.
\end{eqnarray}

\section{\label{app:Z3-gauhing-Z18} Froggatt-Nielsen mechanism in $\mathbb{Z}_3$ gauging of $\mathbb{Z}_{N}$ symmetry}

We first recall the construction of the $\mathbb{Z}_3$ gauging of the $\mathbb{Z}_N$ symmetry~\cite{Dong:2025jra,Qu:2026omn}. Let $a$ and $b$ be the generators of $\mathbb{Z}_N$
and $\mathbb{Z}_3$, respectively, satisfying
\begin{equation}
a^N=b^3=e\,,
\end{equation}
where $e$ denotes the identity element. The generator $b$ acts on $\mathbb{Z}_N$ through an automorphism specified by 
\begin{eqnarray}
\label{eq:Z3-gauging-def}
b^{-1}a b = a^m \,.
\end{eqnarray} 
Consistency with $b^3=e$ requires the integer $m$ to satisfy
\begin{eqnarray}
\label{eq:consist_n3}
m^3 - 1 = 0 ~ (\text{mod}~ N) \,.
\end{eqnarray}
The solutions to the consistency condition therefore include
\begin{eqnarray}
(N, m)=(7, 2), (9, 4), (13, 3), (14, 9), (18, 7), (19,7), (21, 4)  \cdots  \,.
\end{eqnarray}
The action of the $\mathbb{Z}_3$ gauging partitions the elements of $\mathbb{Z}_N$ into equivalence classes. In particular, from Eq.~\eqref{eq:Z3-gauging-def}, we have
\begin{eqnarray}
\label{eq:classes-Z3gauging}\{ a^k, a^{km}, a^{k m^2} \} \equiv [k] \,,
\end{eqnarray}
for each integer $k$. The identity element forms the trivial class $[0] = \{ e \} $. The fusion rules of these equivalence classes are then
\begin{eqnarray}
\label{eq:FR-Z3gauging}[k] \times [l]= [k+l] + [k+lm] + [k+lm^2]\,.
\end{eqnarray} 

Following the same approach, the Froggatt--Nielsen mechanism based on $\mathbb{Z}_2$ gauging of $\mathbb{Z}_N$ or $U(1)$, introduced in section~\ref{sec:framework-FN}, can be extended to the $\mathbb{Z}_3$ gauging of the $\mathbb{Z}_N$ symmetry. When the flavon $S$ is associated with a nontrivial equivalence class $[k_S]\neq[0]$, its complex conjugate $S^{*}$ belongs to the class $[-k_S]$, which is distinct from $[k_S]$. Therefore, $S$ necessarily has to be a complex scalar field. As demonstrated in Appendix~\ref{app:VEV-H-S}, we expect that there exists a region of parameter space in which both $S$ and the Higgs field $H$ acquire stable and nonzero vacuum expectation values.

In the following, we investigate the $\mathbb{Z}_3$ gauging of $\mathbb{Z}_{18}$ with $m=7$, with the aim of understanding the hierarchical patterns of fermion masses and mixing. In this case, there are ten equivalence classes given by
\begin{eqnarray}
[k] = \{ a^k, a^{6+k}, a^{12 + k} \} \,,~~ [3l] = \{ a^{3l} \}\,,~~ \text{ for }~~ k = 1,2,4,5,~~\text{and}~~ 0\leq l \leq 5 \,.
\end{eqnarray}
The related fusion rules are
 \begin{eqnarray}
 [k] \times [l] = [k+l] + [k+7l] + [k+13l]
\end{eqnarray}
They can be written out explicitly as follows,
\begin{eqnarray}
\nonumber [1] \times [1] &=& 3[2]  \,, ~~~[1] \times [2] = 3[3] + 3[9] + 3[15]  \,,~~~[1] \times [3] = [4]  \,,~~~[1] \times [4] = 3[5]  \,,~~~\\ 
\nonumber [1] \times [5] &=& 3[0] + 3[6] + 3[12]  \,, ~~~[1] \times [6] = [1]  \,,~~~[1] \times [9] = [4]  \,,~~~[1] \times [12] = [1]  \,,~~~\\ 
\nonumber [1] \times [15] &=& [4]  \,, ~~~[2] \times [2] = 3[4]  \,,~~~[2] \times [3] = [5]  \,,~~~[2] \times [4] = 3[0] + 3[6] + 3[12]  \,,~~~\\ 
\nonumber [2] \times [5] &=& 3[1]  \,, ~~~[2] \times [6] = [2]  \,,~~~[2] \times [9] = [5]  \,,~~~[2] \times [12] = [2]  \,,~~~[2] \times [15] = [5]  \,,~~~\\ 
\nonumber [3] \times [3] &=& [6]  \,, ~~~[3] \times [4] = [1]  \,,~~~[3] \times [5] = [2]  \,,~~~[3] \times [6] = [9]  \,,~~~[3] \times [9] = [12]  \,,~~~\\ 
\nonumber [3] \times [12] &=& [15]  \,, ~~~[3] \times [15] = [0]  \,,~~~[4] \times [4] = 3[2]  \,,~~~[4] \times [5] = 3[3] + 3[9] + 3[15]  \,,~~~\\ 
\nonumber [4] \times [6] &=& [4]  \,, ~~~[4] \times [9] = [1]  \,,~~~[4] \times [12] = [4]  \,,~~~[4] \times [15] = [1]  \,,~~~[5] \times [5] = 3[4]  \,,~~~\\ 
\nonumber [5] \times [6] &=& [5]  \,, ~~~[5] \times [9] = [2]  \,,~~~[5] \times [12] = [5]  \,,~~~[5] \times [15] = [2]  \,,~~~[6] \times [6] = [12]  \,,~~~\\ 
\nonumber [6] \times [9] &=& [15]  \,, ~~~[6] \times [12] = [0]  \,,~~~[6] \times [15] = [3]  \,,~~~[9] \times [9] = [0]  \,,~~~[9] \times [12] = [3]  \,,~~~\\ 
~ [9] \times [15] &=& [6]  \,, ~~~[12] \times [12] = [6]  \,,~~~[12] \times [15] = [9]  \,,~~~[15] \times [15] = [12]  \,.
\end{eqnarray}
The scalar field are assigned to be $H\sim [0]$ and $S\sim [1]$ respectively. The quark fields are assigned to be
\begin{eqnarray}
(Q_1, Q_2, Q_3) \sim ([0], [1], [15]) \,,~~~ (u_1, u_2, u_3) \sim ([6], [2], [3]) \,,~~~ (d_1, d_2, d_3) \sim ([6], [1], [1])
\end{eqnarray}
With these assignment the mass matrices have the order of magnitude entries
\begin{eqnarray}
M_u \sim \begin{pmatrix}
\varepsilon^6 ~&~ \varepsilon^4 ~&~ \varepsilon^3   \\
\varepsilon^5 ~&~ \varepsilon^3 ~&~ \varepsilon^2   \\
\varepsilon^3 ~&~ \varepsilon ~&~ 1
\end{pmatrix} v_H \,,~~~
M_d \sim \begin{pmatrix}
\varepsilon^6 ~&~ \varepsilon^5 ~&~ \varepsilon^5   \\
\varepsilon^5 ~&~ \varepsilon^4 ~&~ \varepsilon^4   \\
\varepsilon^3 ~&~ \varepsilon^2 ~&~ \varepsilon^2
\end{pmatrix} v_H \,.
\end{eqnarray}
The above quark mass matrices have the same form as those in Eq.~\eqref{eq:Mu-Md-N12}. The magnitude of quark masses and the mixing parameters are given by Eq.~\eqref{eq:N12_quark_para}.

Under the action of $\mathbb{Z}_3$ gauging of $\mathbb{Z}_{18}$ symmetry, the lepton fields are assigned to 
\begin{eqnarray}
\ell \sim ([6], [1], [1])\,,~~~ E_R \sim ([6], [1], [9])\,,~~~\nu_R \sim ([4], [5], [9])\,.
\end{eqnarray}
The charged lepton mass and neutrino mass matrices are
\begin{eqnarray}
M_e &\sim & \begin{pmatrix}
\varepsilon^6 ~&~ \varepsilon^5 ~&~ \varepsilon^3   \\
\varepsilon^5 ~&~ \varepsilon^4 ~&~ \varepsilon^2   \\
\varepsilon^5 ~&~ \varepsilon^4 ~&~ \varepsilon^2
\end{pmatrix} v_H  \,.
\end{eqnarray}
The neutrino mass matrices are
\begin{eqnarray}\nonumber 
M_D &\sim &  \begin{pmatrix}
\varepsilon^2 ~&~ \varepsilon ~&~ \varepsilon^3   \\
\varepsilon ~&~ 1 ~&~ \varepsilon^2   \\
\varepsilon ~&~ 1 ~&~ \varepsilon^2
\end{pmatrix} v_H \,,~~~   
M_N \sim \begin{pmatrix}
\varepsilon^4 ~&~ \varepsilon^3 ~&~ \varepsilon^5   \\
\varepsilon^3 ~&~ \varepsilon^2 ~&~ \varepsilon^4   \\
\varepsilon^5 ~&~ \varepsilon^4 ~&~ 1
\end{pmatrix} M \,,~~~   \\
M_N^{-1} &\sim & \begin{pmatrix}
1 ~&~ \varepsilon ~&~ \varepsilon^5   \\
\varepsilon ~&~ \varepsilon^2 ~&~ \varepsilon^6   \\
\varepsilon^5 ~&~ \varepsilon^6 ~&~ \varepsilon^4
\end{pmatrix} \dfrac{1}{\varepsilon^4 M} \,,~~~
M_\nu \sim \begin{pmatrix}
\varepsilon^2 ~&~ \varepsilon ~&~ \varepsilon   \\
\varepsilon ~&~ 1 ~&~ 1   \\
\varepsilon ~&~ 1 ~&~ 1
\end{pmatrix} \dfrac{v_H^2}{\varepsilon^2 M} \,.
\end{eqnarray}
Compared with the $\mathbb{Z}_2$ gauging of $\mathbb{Z}_N$ symmetry, the additional suppressions $\varepsilon^4$ and $\varepsilon^2$ in the (11) and (22) entries of $M_N$, respectively, naturally generate a hierarchical right-handed neutrino mass spectrum,  $M_1: M_2: M_3\approx \varepsilon^4: \varepsilon^2: 1$. The order of magnitudes predictions for the charged lepton masses and neutrino mixing angels parameters,
\begin{eqnarray}
\nonumber&& m_e\sim\varepsilon^6 v_H\,,~~ m_\mu\sim\varepsilon^4 v_H\,,~~m_{\tau}\sim\varepsilon^2 v_H\,, \\
\nonumber&&\sin\theta_{12}\sim\varepsilon\,,~~\sin\theta_{13}\sim\varepsilon \,,~~\sin\theta_{23}\sim1 \,,  \\
&& m_1\sim \varepsilon^6\dfrac{v_H^2}{M}\,,~m_2 \sim \dfrac{v_H^2}{\varepsilon^2 M}\,,~ m_3 \sim \dfrac{v_H^2}{\varepsilon^2 M}\,.
\end{eqnarray}

\end{appendix}

\providecommand{\href}[2]{#2}\begingroup\raggedright\endgroup


\begin{thebibliography}{10}

\bibitem{ParticleDataGroup:2026mpi}
{\bfseries Particle Data Group} Collaboration, F.~Takahashi {\em et~al.},
  ``{Review of Particle Physics},''
  \href{http://dx.doi.org/10.1142/s0217751x26300115}{{\em Int. J. Mod. Phys. A}
  {\bfseries 41} no.~22, (2026) 2630011}.

\bibitem{Froggatt:1978nt}
C.~D. Froggatt and H.~B. Nielsen, ``{Hierarchy of Quark Masses, Cabibbo Angles
  and CP Violation},''
  \href{http://dx.doi.org/10.1016/0550-3213(79)90316-X}{{\em Nucl. Phys. B}
  {\bfseries 147} (1979) 277--298}.

\bibitem{Weinberg:1977hb}
S.~Weinberg, ``{The Problem of Mass},''
  \href{http://dx.doi.org/10.1111/j.2164-0947.1977.tb02958.x}{{\em Trans. New
  York Acad. Sci.} {\bfseries 38} (1977) 185--201}.

\bibitem{Fritzsch:1977za}
H.~Fritzsch, ``{Calculating the Cabibbo Angle},''
  \href{http://dx.doi.org/10.1016/0370-2693(77)90408-7}{{\em Phys. Lett. B}
  {\bfseries 70} (1977) 436--440}.

\bibitem{Wilczek:1977uh}
F.~Wilczek and A.~Zee, ``{Discrete Flavor Symmetries and a Formula for the
  Cabibbo Angle},'' \href{http://dx.doi.org/10.1016/0370-2693(77)90403-8}{{\em
  Phys. Lett. B} {\bfseries 70} (1977) 418}. [Erratum: Phys.Lett.B 72, 504
  (1978)].

\bibitem{Frampton:2002yf}
P.~H. Frampton, S.~L. Glashow, and D.~Marfatia, ``{Zeroes of the neutrino mass
  matrix},'' \href{http://dx.doi.org/10.1016/S0370-2693(02)01817-8}{{\em Phys.
  Lett. B} {\bfseries 536} (2002) 79--82},
  \href{http://arxiv.org/abs/hep-ph/0201008}{{\ttfamily arXiv:hep-ph/0201008}}.

\bibitem{Xing:2002ta}
Z.-z. Xing, ``{Texture zeros and Majorana phases of the neutrino mass
  matrix},'' \href{http://dx.doi.org/10.1016/S0370-2693(02)01354-0}{{\em Phys.
  Lett. B} {\bfseries 530} (2002) 159--166},
  \href{http://arxiv.org/abs/hep-ph/0201151}{{\ttfamily arXiv:hep-ph/0201151}}.

\bibitem{Gaiotto:2014kfa}
D.~Gaiotto, A.~Kapustin, N.~Seiberg, and B.~Willett, ``{Generalized Global
  Symmetries},'' \href{http://dx.doi.org/10.1007/JHEP02(2015)172}{{\em JHEP}
  {\bfseries 02} (2015) 172}, \href{http://arxiv.org/abs/1412.5148}{{\ttfamily
  arXiv:1412.5148 [hep-th]}}.

\bibitem{Cordova:2022ruw}
C.~Cordova, T.~T. Dumitrescu, K.~Intriligator, and S.-H. Shao, ``{Snowmass
  White Paper: Generalized Symmetries in Quantum Field Theory and Beyond},'' in
  {\em {Snowmass 2021}}.
\newblock 5, 2022.
\newblock \href{http://arxiv.org/abs/2205.09545}{{\ttfamily arXiv:2205.09545
  [hep-th]}}.

\bibitem{Schafer-Nameki:2023jdn}
S.~Schafer-Nameki, ``{ICTP lectures on (non-)invertible generalized
  symmetries},'' \href{http://dx.doi.org/10.1016/j.physrep.2024.01.007}{{\em
  Phys. Rept.} {\bfseries 1063} (2024) 1--55},
  \href{http://arxiv.org/abs/2305.18296}{{\ttfamily arXiv:2305.18296
  [hep-th]}}.

\bibitem{Brennan:2023mmt}
T.~D. Brennan and S.~Hong, ``{Introduction to Generalized Global Symmetries in
  QFT and Particle Physics},''
  \href{http://arxiv.org/abs/2306.00912}{{\ttfamily arXiv:2306.00912
  [hep-ph]}}.

\bibitem{Bhardwaj:2023kri}
L.~Bhardwaj, L.~E. Bottini, L.~Fraser-Taliente, L.~Gladden, D.~S.~W. Gould,
  A.~Platschorre, and H.~Tillim, ``{Lectures on generalized symmetries},''
  \href{http://dx.doi.org/10.1016/j.physrep.2023.11.002}{{\em Phys. Rept.}
  {\bfseries 1051} (2024) 1--87},
  \href{http://arxiv.org/abs/2307.07547}{{\ttfamily arXiv:2307.07547
  [hep-th]}}.

\bibitem{Shao:2023gho}
S.-H. Shao, ``{What's Done Cannot Be Undone: TASI Lectures on Non-Invertible
  Symmetries},'' \href{http://arxiv.org/abs/2308.00747}{{\ttfamily
  arXiv:2308.00747 [hep-th]}}.

\bibitem{Kobayashi:2024cvp}
T.~Kobayashi, H.~Otsuka, and M.~Tanimoto, ``{Yukawa textures from
  non-invertible symmetries},''
  \href{http://dx.doi.org/10.1007/JHEP12(2024)117}{{\em JHEP} {\bfseries 12}
  (2024) 117}, \href{http://arxiv.org/abs/2409.05270}{{\ttfamily
  arXiv:2409.05270 [hep-ph]}}.

\bibitem{Kobayashi:2025znw}
T.~Kobayashi, Y.~Nishioka, H.~Otsuka, and M.~Tanimoto, ``{More about quark
  Yukawa textures from selection rules without group actions},''
  \href{http://dx.doi.org/10.1007/JHEP05(2025)177}{{\em JHEP} {\bfseries 05}
  (2025) 177}, \href{http://arxiv.org/abs/2503.09966}{{\ttfamily
  arXiv:2503.09966 [hep-ph]}}.

\bibitem{Kobayashi:2025ldi}
T.~Kobayashi, H.~Otsuka, M.~Tanimoto, and H.~Uchida, ``{Lepton mass textures
  from non-invertible multiplication rules},''
  \href{http://arxiv.org/abs/2505.07262}{{\ttfamily arXiv:2505.07262
  [hep-ph]}}.

\bibitem{Kobayashi:2025cwx}
T.~Kobayashi, H.~Okada, and H.~Otsuka, ``{Radiative neutrino mass models from
  non-invertible selection rules},''
  \href{http://arxiv.org/abs/2505.14878}{{\ttfamily arXiv:2505.14878
  [hep-ph]}}.

\bibitem{Chen:2025awz}
J.~Chen, C.-Q. Geng, H.~Okada, and J.-J. Wu, ``{A radiative lepton model in a
  non-invertible fusion rule},''
  \href{http://arxiv.org/abs/2507.11951}{{\ttfamily arXiv:2507.11951
  [hep-ph]}}.

\bibitem{Qu:2026omn}
B.-Y. Qu, Z.~Jiang, and G.-J. Ding, ``{Two-zero textures of the Majorana
  neutrino mass matrix from {\ensuremath{\mathbb{Z}}}$_{3}$ gauging of
  {\ensuremath{\mathbb{Z}}}$_{N}$ non-invertible symmetry},''
  \href{http://dx.doi.org/10.1007/JHEP07(2026)245}{{\em JHEP} {\bfseries 07}
  (2026) 245}, \href{http://arxiv.org/abs/2602.24214}{{\ttfamily
  arXiv:2602.24214 [hep-ph]}}.

\bibitem{Jiang:2025psz}
Z.~Jiang, B.-Y. Qu, and G.-J. Ding, ``{Texture-zeros in minimal seesaw from
  noninvertible symmetry fusion rules},''
  \href{http://dx.doi.org/10.1103/d29s-cw34}{{\em Phys. Rev. D} {\bfseries 112}
  no.~11, (2025) 115029}, \href{http://arxiv.org/abs/2510.07236}{{\ttfamily
  arXiv:2510.07236 [hep-ph]}}.

\bibitem{Liang:2025dkm}
Q.~Liang and T.~T. Yanagida, ``{Non-invertible symmetry as an axion-less
  solution to the strong CP problem},''
  \href{http://dx.doi.org/10.1016/j.physletb.2025.139706}{{\em Phys. Lett. B}
  {\bfseries 868} (2025) 139706},
  \href{http://arxiv.org/abs/2505.05142}{{\ttfamily arXiv:2505.05142
  [hep-ph]}}.

\bibitem{Kobayashi:2025thd}
T.~Kobayashi, H.~Otsuka, and T.~T. Yanagida, ``{Non-invertible Symmetry as a
  Solution to the Strong CP Problem in a GUT-inspired Standard Model},''
  \href{http://arxiv.org/abs/2508.12287}{{\ttfamily arXiv:2508.12287
  [hep-ph]}}.

\bibitem{Kobayashi:2025rpx}
T.~Kobayashi, H.~Otsuka, M.~Tanimoto, and T.~T. Yanagida, ``{GUT-motivated
  non-invertible symmetry as a solution to the strong CP problem and the
  neutrino CP-violating phase},''
  \href{http://arxiv.org/abs/2510.01680}{{\ttfamily arXiv:2510.01680
  [hep-ph]}}.

\bibitem{King:1999mb}
S.~F. King, ``{Large mixing angle MSW and atmospheric neutrinos from single
  right-handed neutrino dominance and U(1) family symmetry},''
  \href{http://dx.doi.org/10.1016/S0550-3213(00)00109-7}{{\em Nucl. Phys. B}
  {\bfseries 576} (2000) 85--105},
  \href{http://arxiv.org/abs/hep-ph/9912492}{{\ttfamily arXiv:hep-ph/9912492}}.

\bibitem{Frampton:2002qc}
P.~H. Frampton, S.~L. Glashow, and T.~Yanagida, ``{Cosmological sign of
  neutrino CP violation},''
  \href{http://dx.doi.org/10.1016/S0370-2693(02)02853-8}{{\em Phys. Lett. B}
  {\bfseries 548} (2002) 119--121},
  \href{http://arxiv.org/abs/hep-ph/0208157}{{\ttfamily arXiv:hep-ph/0208157}}.

\bibitem{Bartsch:2022mpm}
T.~Bartsch, M.~Bullimore, A.~E.~V. Ferrari, and J.~Pearson, ``{Non-invertible
  symmetries and higher representation theory I},''
  \href{http://dx.doi.org/10.21468/SciPostPhys.17.1.015}{{\em SciPost Phys.}
  {\bfseries 17} no.~1, (2024) 015},
  \href{http://arxiv.org/abs/2208.05993}{{\ttfamily arXiv:2208.05993
  [hep-th]}}.

\bibitem{Ibanez:1994ig}
L.~E. Ibanez and G.~G. Ross, ``{Fermion masses and mixing angles from gauge
  symmetries},'' \href{http://dx.doi.org/10.1016/0370-2693(94)90865-6}{{\em
  Phys. Lett. B} {\bfseries 332} (1994) 100--110},
  \href{http://arxiv.org/abs/hep-ph/9403338}{{\ttfamily arXiv:hep-ph/9403338}}.

\bibitem{Lee00}
J.~M. Lee, ``Introduction to smooth manifolds,''.

\bibitem{Esteban:2024eli}
I.~Esteban, M.~C. Gonzalez-Garcia, M.~Maltoni, I.~Martinez-Soler, J.~P.
  Pinheiro, and T.~Schwetz, ``{NuFit-6.0: updated global analysis of
  three-flavor neutrino oscillations},''
  \href{http://dx.doi.org/10.1007/JHEP12(2024)216}{{\em JHEP} {\bfseries 12}
  (2024) 216}, \href{http://arxiv.org/abs/2410.05380}{{\ttfamily
  arXiv:2410.05380 [hep-ph]}}.

\bibitem{Cornella:2023zme}
C.~Cornella, D.~Curtin, E.~T. Neil, and J.~O. Thompson, ``{Mapping and probing
  Froggatt-Nielsen solutions to the quark flavor puzzle},''
  \href{http://dx.doi.org/10.1103/PhysRevD.111.015042}{{\em Phys. Rev. D}
  {\bfseries 111} no.~1, (2025) 015042},
  \href{http://arxiv.org/abs/2306.08026}{{\ttfamily arXiv:2306.08026
  [hep-ph]}}.

\bibitem{Antusch:2025fpm}
S.~Antusch, K.~Hinze, and S.~Saad, ``{Updated running quark and lepton
  parameters at various scales},''
  \href{http://dx.doi.org/10.1103/fdcc-ycph}{{\em Phys. Rev. D} {\bfseries 113}
  no.~9, (2026) 095011}, \href{http://arxiv.org/abs/2510.01312}{{\ttfamily
  arXiv:2510.01312 [hep-ph]}}.

\bibitem{Cornella:2024jaw}
C.~Cornella, D.~Curtin, G.~Krnjaic, and M.~Mellors, ``{Testing the
  Froggatt-Nielsen mechanism with lepton flavor and number violating
  processes},'' \href{http://dx.doi.org/10.1103/c2ws-hx4h}{{\em Phys. Rev. D}
  {\bfseries 112} no.~11, (2025) 115010},
  \href{http://arxiv.org/abs/2501.00629}{{\ttfamily arXiv:2501.00629
  [hep-ph]}}.

\bibitem{Gatto:1968ss}
R.~Gatto, G.~Sartori, and M.~Tonin, ``{Weak Selfmasses, Cabibbo Angle, and
  Broken SU(2) x SU(2)},''
  \href{http://dx.doi.org/10.1016/0370-2693(68)90150-0}{{\em Phys. Lett. B}
  {\bfseries 28} (1968) 128--130}.

\bibitem{Ponce:2011qp}
W.~A. Ponce and R.~H. Benavides, ``{Texture Zeros for the Standard Model Quark
  Mass Matrices},''
  \href{http://dx.doi.org/10.1140/epjc/s10052-011-1641-2}{{\em Eur. Phys. J. C}
  {\bfseries 71} (2011) 1641}, \href{http://arxiv.org/abs/1104.5475}{{\ttfamily
  arXiv:1104.5475 [hep-ph]}}.

\bibitem{Grimus:2004hf}
W.~Grimus, A.~S. Joshipura, L.~Lavoura, and M.~Tanimoto, ``{Symmetry
  realization of texture zeros},''
  \href{http://dx.doi.org/10.1140/epjc/s2004-01896-y}{{\em Eur. Phys. J. C}
  {\bfseries 36} (2004) 227--232},
  \href{http://arxiv.org/abs/hep-ph/0405016}{{\ttfamily arXiv:hep-ph/0405016}}.

\bibitem{Linster:2018avp}
M.~Linster and R.~Ziegler, ``{A Realistic $U(2)$ Model of Flavor},''
  \href{http://dx.doi.org/10.1007/JHEP08(2018)058}{{\em JHEP} {\bfseries 08}
  (2018) 058}, \href{http://arxiv.org/abs/1805.07341}{{\ttfamily
  arXiv:1805.07341 [hep-ph]}}.

\bibitem{Arkani-Hamed:2026uos}
N.~Arkani-Hamed, C.~Figueiredo, L.~J. Hall, and C.~A. Manzari, ``{The Very
  Nearly Right Theory of Flavor},''
  \href{http://arxiv.org/abs/2607.27315}{{\ttfamily arXiv:2607.27315
  [hep-ph]}}.

\bibitem{King:2002nf}
S.~F. King, ``{Constructing the large mixing angle MNS matrix in seesaw models
  with right-handed neutrino dominance},''
  \href{http://dx.doi.org/10.1088/1126-6708/2002/09/011}{{\em JHEP} {\bfseries
  09} (2002) 011}, \href{http://arxiv.org/abs/hep-ph/0204360}{{\ttfamily
  arXiv:hep-ph/0204360}}.

\bibitem{KamLAND-Zen:2024eml}
{\bfseries KamLAND-Zen} Collaboration, S.~Abe {\em et~al.}, ``{Search for
  Majorana Neutrinos with the Complete KamLAND-Zen Dataset},''
  \href{http://dx.doi.org/10.1103/jkf6-48j8}{{\em Phys. Rev. Lett.} {\bfseries
  135} no.~26, (2025) 262501},
  \href{http://arxiv.org/abs/2406.11438}{{\ttfamily arXiv:2406.11438
  [hep-ex]}}.

\bibitem{LEGEND:2021bnm}
{\bfseries LEGEND} Collaboration, N.~Abgrall {\em et~al.}, ``{The Large
  Enriched Germanium Experiment for Neutrinoless $\beta\beta$ Decay}:
  {LEGEND-1000 Preconceptual Design Report},''
  \href{http://arxiv.org/abs/2107.11462}{{\ttfamily arXiv:2107.11462
  [physics.ins-det]}}.

\bibitem{nEXO:2021ujk}
{\bfseries nEXO} Collaboration, G.~Adhikari {\em et~al.}, ``{nEXO: neutrinoless
  double beta decay search beyond 10$^{28}$ year half-life sensitivity},''
  \href{http://dx.doi.org/10.1088/1361-6471/ac3631}{{\em J. Phys. G} {\bfseries
  49} no.~1, (2022) 015104}, \href{http://arxiv.org/abs/2106.16243}{{\ttfamily
  arXiv:2106.16243 [nucl-ex]}}.

\bibitem{KATRIN:2024cdt}
{\bfseries KATRIN} Collaboration, M.~Aker {\em et~al.}, ``{Direct neutrino-mass
  measurement based on 259 days of KATRIN data},''
  \href{http://dx.doi.org/10.1126/science.adq9592}{{\em Science} {\bfseries
  388} no.~6743, (2025) adq9592},
  \href{http://arxiv.org/abs/2406.13516}{{\ttfamily arXiv:2406.13516
  [nucl-ex]}}.

\bibitem{KATRIN:2021dfa}
{\bfseries KATRIN} Collaboration, M.~Aker {\em et~al.}, ``{The design,
  construction, and commissioning of the KATRIN experiment},''
  \href{http://dx.doi.org/10.1088/1748-0221/16/08/T08015}{{\em JINST}
  {\bfseries 16} no.~08, (2021) T08015},
  \href{http://arxiv.org/abs/2103.04755}{{\ttfamily arXiv:2103.04755
  [physics.ins-det]}}.

\bibitem{Project8:2022wqh}
{\bfseries Project 8} Collaboration, A.~A. Esfahani {\em et~al.}, ``{The
  Project 8 Neutrino Mass Experiment},'' in {\em {Snowmass 2021}}.
\newblock 3, 2022.
\newblock \href{http://arxiv.org/abs/2203.07349}{{\ttfamily arXiv:2203.07349
  [nucl-ex]}}.

\bibitem{Planck:2018vyg}
{\bfseries Planck} Collaboration, N.~Aghanim {\em et~al.}, ``{Planck 2018
  results. VI. Cosmological parameters},''
  \href{http://dx.doi.org/10.1051/0004-6361/201833910}{{\em Astron. Astrophys.}
  {\bfseries 641} (2020) A6}, \href{http://arxiv.org/abs/1807.06209}{{\ttfamily
  arXiv:1807.06209 [astro-ph.CO]}}. [Erratum: Astron.Astrophys. 652, C4
  (2021)].

\bibitem{Euclid:2024imf}
{\bfseries Euclid} Collaboration, M.~Archidiacono {\em et~al.}, ``{Euclid
  preparation - LIV. Sensitivity to neutrino parameters},''
  \href{http://dx.doi.org/10.1051/0004-6361/202450859}{{\em Astron. Astrophys.}
  {\bfseries 693} (2025) A58},
  \href{http://arxiv.org/abs/2405.06047}{{\ttfamily arXiv:2405.06047
  [astro-ph.CO]}}.

\bibitem{Feroz:2007kg}
F.~Feroz and M.~P. Hobson, ``{Multimodal nested sampling: an efficient and
  robust alternative to MCMC methods for astronomical data analysis},''
  \href{http://dx.doi.org/10.1111/j.1365-2966.2007.12353.x}{{\em Mon. Not. Roy.
  Astron. Soc.} {\bfseries 384} (2008) 449},
  \href{http://arxiv.org/abs/0704.3704}{{\ttfamily arXiv:0704.3704
  [astro-ph]}}.

\bibitem{Feroz:2008xx}
F.~Feroz, M.~P. Hobson, and M.~Bridges, ``{MultiNest: an efficient and robust
  Bayesian inference tool for cosmology and particle physics},''
  \href{http://dx.doi.org/10.1111/j.1365-2966.2009.14548.x}{{\em Mon. Not. Roy.
  Astron. Soc.} {\bfseries 398} (2009) 1601--1614},
  \href{http://arxiv.org/abs/0809.3437}{{\ttfamily arXiv:0809.3437
  [astro-ph]}}.

\bibitem{Otsuka:2026fpy}
H.~Otsuka and R.~Yokokura, ``{Non-Invertible Froggatt-Nielsen Symmetry},''
  \href{http://arxiv.org/abs/2610.21688}{{\ttfamily arXiv:2610.21688
  [hep-ph]}}.

\bibitem{Burgess:2000yq}
C.~P. Burgess, M.~Pospelov, and T.~ter Veldhuis, ``{The Minimal model of
  nonbaryonic dark matter: A Singlet scalar},''
  \href{http://dx.doi.org/10.1016/S0550-3213(01)00513-2}{{\em Nucl. Phys. B}
  {\bfseries 619} (2001) 709--728},
  \href{http://arxiv.org/abs/hep-ph/0011335}{{\ttfamily arXiv:hep-ph/0011335}}.

\bibitem{Baldes:2016gaf}
I.~Baldes, T.~Konstandin, and G.~Servant, ``{Flavor Cosmology: Dynamical
  Yukawas in the Froggatt-Nielsen Mechanism},''
  \href{http://dx.doi.org/10.1007/JHEP12(2016)073}{{\em JHEP} {\bfseries 12}
  (2016) 073}, \href{http://arxiv.org/abs/1608.03254}{{\ttfamily
  arXiv:1608.03254 [hep-ph]}}.

\bibitem{Kannike:2012pe}
K.~Kannike, ``{Vacuum Stability Conditions From Copositivity Criteria},''
  \href{http://dx.doi.org/10.1140/epjc/s10052-012-2093-z}{{\em Eur. Phys. J. C}
  {\bfseries 72} (2012) 2093}, \href{http://arxiv.org/abs/1205.3781}{{\ttfamily
  arXiv:1205.3781 [hep-ph]}}.

\bibitem{Dong:2025jra}
J.~Dong, T.~Jeric, T.~Kobayashi, R.~Nishida, and H.~Otsuka, ``{Discrete gauging
  and noninvertible selection rules},''
  \href{http://dx.doi.org/10.1103/nsvv-l2dy}{{\em Phys. Rev. D} {\bfseries 113}
  no.~5, (2026) 056028}, \href{http://arxiv.org/abs/2507.02375}{{\ttfamily
  arXiv:2507.02375 [hep-th]}}.

\end{thebibliography}
\end{document}